\documentstyle[10pt,epsf,epsfig,dp_delphititle]{dp_delphi}
\makeindex
\def\DpPaperGroup{EP}
\def\DpPaperRef{2000-021}
\def\DpDate{31 January 2000}
\def\DpAuthors{DELPHI Collaboration}
\def\DpSubmit{(Eur. Phys. J. C17(2000)53)}
\def\DpTitle{{\bf Photon Events with Missing Energy at 
              \boldmath $\sqrt{s} =$183 to 189 GeV }}
\def\DpComment{ }
\def\DpEMail{ } 

\newcommand {\ee} {\rm{e}^+\rm{e}^-}
\newcommand {\MZ} {\rm{m_Z}}
\newcommand {\eenng} {e^+e^-\rightarrow \nu\bar{\nu}\gamma}
\newcommand {\eeGG} {e^+e^-\rightarrow\tilde{G}\tilde{G}\gamma}
\newcommand {\eeGGg}
{e^+e^-\rightarrow\tilde{G}\tilde{\chi}^0_1\rightarrow\tilde{G}\tilde{G}\gamma}
\newcommand {\eeGgGg}
{e^+e^-\rightarrow
\tilde{\chi}^0_1\tilde{\chi}^0_1\rightarrow\tilde{G}\gamma\tilde{G}\gamma}

\newcommand {\eeXgXg}
{e^+e^-\rightarrow\tilde{\chi}^0_2\tilde{\chi}^0_2\rightarrow\tilde{\chi}^0_1
\gamma\tilde{\chi}^0_1\gamma}

\def\gt{\raisebox{0.2ex}{$>$}}
\begin{document}
\makeatletter
\newcount\@tempcntc
\def\@citex[#1]#2{\if@filesw\immediate\write\@auxout{\string\citation{#2}}\fi
  \@tempcnta\z@\@tempcntb\m@ne\def\@citea{}\@cite{\@for\@citeb:=#2\do
    {\@ifundefined
       {b@\@citeb}{\@citeo\@tempcntb\m@ne\@citea\def\@citea{,}{\bf ?}\@warning
       {Citation `\@citeb' on page \thepage \space undefined}}%
    {\setbox\z@\hbox{\global\@tempcntc0\csname b@\@citeb\endcsname\relax}%
     \ifnum\@tempcntc=\z@ \@citeo\@tempcntb\m@ne
       \@citea\def\@citea{,}\hbox{\csname b@\@citeb\endcsname}%
     \else
      \advance\@tempcntb\@ne
      \ifnum\@tempcntb=\@tempcntc
      \else\advance\@tempcntb\m@ne\@citeo
      \@tempcnta\@tempcntc\@tempcntb\@tempcntc\fi\fi}}\@citeo}{#1}}
\def\@citeo{\ifnum\@tempcnta>\@tempcntb\else\@citea\def\@citea{,}%
  \ifnum\@tempcnta=\@tempcntb\the\@tempcnta\else
   {\advance\@tempcnta\@ne\ifnum\@tempcnta=\@tempcntb \else \def\@citea{--}\fi
    \advance\@tempcnta\m@ne\the\@tempcnta\@citea\the\@tempcntb}\fi\fi}
 
\makeatother
\begin{titlepage}
\pagenumbering{roman}
\CERNpreprint{\DpPaperGroup}{\DpPaperRef} 
\date{{\small\DpDate}} 
\title{\DpTitle} 
\address{\DpAuthors} 
\begin{shortabs} 
\noindent
%
\noindent

The production of single photons has been studied
in the reaction $e^+e^-\rightarrow\gamma+invisible~particles$ at
centre-of-mass energies of 183~GeV and 189~GeV. 
A previously published analysis of events with multi-photon
final states accompanied by missing energy has been updated with 189~GeV data.
The data were collected with the DELPHI detector and correspond to integrated
luminosities of about 51~pb$^{-1}$ and 158~pb$^{-1}$ at the two energies.
The number of light neutrino families is measured to be
$2.84\pm0.15(stat)\pm0.14(syst)$. The absence of
an excess of events beyond that expected from Standard Model
processes is used to set limits on new physics
as described by supersymmetric and composite models.
A limit on the gravitational scale is also determined.
\end{shortabs}
\vfill
\begin{center}
\DpSubmit \ \\ 
\DpComment \ \\
\DpEMail \ \\
\end{center}
\vfill
\clearpage
\headsep 10.0pt
\addtolength{\textheight}{10mm}
\addtolength{\footskip}{-5mm}
\begingroup
%
\newcommand{\DpName}[2]{\hbox{#1$^{\ref{#2}}$},\hfill}
\newcommand{\DpNameTwo}[3]{\hbox{#1$^{\ref{#2},\ref{#3}}$},\hfill}
\newcommand{\DpNameThree}[4]{\hbox{#1$^{\ref{#2},\ref{#3},\ref{#4}}$},\hfill}
\newskip\Bigfill \Bigfill = 0pt plus 1000fill
\newcommand{\DpNameLast}[2]{\hbox{#1$^{\ref{#2}}$}\hspace{\Bigfill}}
%
\footnotesize
\noindent
\DpName{P.Abreu}{LIP}
\DpName{W.Adam}{VIENNA}
\DpName{T.Adye}{RAL}
\DpName{P.Adzic}{DEMOKRITOS}
\DpName{I.Ajinenko}{SERPUKHOV}
\DpName{Z.Albrecht}{KARLSRUHE}
\DpName{T.Alderweireld}{AIM}
\DpName{G.D.Alekseev}{JINR}
\DpName{R.Alemany}{VALENCIA}
\DpName{T.Allmendinger}{KARLSRUHE}
% ...
\normalsize
\endgroup
\titlefoot{Department of Physics and Astronomy, Iowa State
     University, Ames IA 50011-3160, USA
    \label{AMES}}
\titlefoot{Physics Department, Univ. Instelling Antwerpen,
     Universiteitsplein 1, B-2610 Antwerpen, Belgium \\
     \indent~~and IIHE, ULB-VUB,
     Pleinlaan 2, B-1050 Brussels, Belgium \\
     \indent~~and Facult\'e des Sciences,
     Univ. de l'Etat Mons, Av. Maistriau 19, B-7000 Mons, Belgium
    \label{AIM}}
\titlefoot{Physics Laboratory, University of Athens, Solonos Str.
     104, GR-10680 Athens, Greece
    \label{ATHENS}}
\titlefoot{Department of Physics, University of Bergen,
     All\'egaten 55, NO-5007 Bergen, Norway
    \label{BERGEN}}
\titlefoot{Dipartimento di Fisica, Universit\`a di Bologna and INFN,
     Via Irnerio 46, IT-40126 Bologna, Italy
    \label{BOLOGNA}}
\titlefoot{Centro Brasileiro de Pesquisas F\'{\i}sicas, rua Xavier Sigaud 150,
     BR-22290 Rio de Janeiro, Brazil \\
     \indent~~and Depto. de F\'{\i}sica, Pont. Univ. Cat\'olica,
     C.P. 38071 BR-22453 Rio de Janeiro, Brazil \\
     \indent~~and Inst. de F\'{\i}sica, Univ. Estadual do Rio de Janeiro,
     rua S\~{a}o Francisco Xavier 524, Rio de Janeiro, Brazil
    \label{BRASIL}}
\titlefoot{Comenius University, Faculty of Mathematics and Physics,
     Mlynska Dolina, SK-84215 Bratislava, Slovakia
    \label{BRATISLAVA}}
\titlefoot{Coll\`ege de France, Lab. de Physique Corpusculaire, IN2P3-CNRS,
     FR-75231 Paris Cedex 05, France
    \label{CDF}}
\titlefoot{CERN, CH-1211 Geneva 23, Switzerland
    \label{CERN}}
\titlefoot{Institut de Recherches Subatomiques, IN2P3 - CNRS/ULP - BP20,
     FR-67037 Strasbourg Cedex, France
    \label{CRN}}
\titlefoot{Now at DESY-Zeuthen, Platanenallee 6, D-15735 Zeuthen, Germany
    \label{DESY}}
\titlefoot{Institute of Nuclear Physics, N.C.S.R. Demokritos,
     P.O. Box 60228, GR-15310 Athens, Greece
    \label{DEMOKRITOS}}
\titlefoot{FZU, Inst. of Phys. of the C.A.S. High Energy Physics Division,
     Na Slovance 2, CZ-180 40, Praha 8, Czech Republic
    \label{FZU}}
\titlefoot{Dipartimento di Fisica, Universit\`a di Genova and INFN,
     Via Dodecaneso 33, IT-16146 Genova, Italy
    \label{GENOVA}}
\titlefoot{Institut des Sciences Nucl\'eaires, IN2P3-CNRS, Universit\'e
     de Grenoble 1, FR-38026 Grenoble Cedex, France
    \label{GRENOBLE}}
\titlefoot{Helsinki Institute of Physics, HIP,
     P.O. Box 9, FI-00014 Helsinki, Finland
    \label{HELSINKI}}
\titlefoot{Joint Institute for Nuclear Research, Dubna, Head Post
     Office, P.O. Box 79, RU-101 000 Moscow, Russian Federation
    \label{JINR}}
\titlefoot{Institut f\"ur Experimentelle Kernphysik,
     Universit\"at Karlsruhe, Postfach 6980, DE-76128 Karlsruhe,
     Germany
    \label{KARLSRUHE}}
\titlefoot{Institute of Nuclear Physics and University of Mining and Metalurgy,
     Ul. Kawiory 26a, PL-30055 Krakow, Poland
    \label{KRAKOW}}
\titlefoot{Universit\'e de Paris-Sud, Lab. de l'Acc\'el\'erateur
     Lin\'eaire, IN2P3-CNRS, B\^{a}t. 200, FR-91405 Orsay Cedex, France
    \label{LAL}}
\titlefoot{School of Physics and Chemistry, University of Lancaster,
     Lancaster LA1 4YB, UK
    \label{LANCASTER}}
\titlefoot{LIP, IST, FCUL - Av. Elias Garcia, 14-$1^{o}$,
     PT-1000 Lisboa Codex, Portugal
    \label{LIP}}
\titlefoot{Department of Physics, University of Liverpool, P.O.
     Box 147, Liverpool L69 3BX, UK
    \label{LIVERPOOL}}
\titlefoot{LPNHE, IN2P3-CNRS, Univ.~Paris VI et VII, Tour 33 (RdC),
     4 place Jussieu, FR-75252 Paris Cedex 05, France
    \label{LPNHE}}
\titlefoot{Department of Physics, University of Lund,
     S\"olvegatan 14, SE-223 63 Lund, Sweden
    \label{LUND}}
\titlefoot{Universit\'e Claude Bernard de Lyon, IPNL, IN2P3-CNRS,
     FR-69622 Villeurbanne Cedex, France
    \label{LYON}}
\titlefoot{Univ. d'Aix - Marseille II - CPP, IN2P3-CNRS,
     FR-13288 Marseille Cedex 09, France
    \label{MARSEILLE}}
\titlefoot{Dipartimento di Fisica, Universit\`a di Milano and INFN-MILANO,
     Via Celoria 16, IT-20133 Milan, Italy
    \label{MILANO}}
\titlefoot{Dipartimento di Fisica, Univ. di Milano-Bicocca and
     INFN-MILANO, Piazza delle Scienze 2, IT-20126 Milan, Italy
    \label{MILANO2}}
\titlefoot{Niels Bohr Institute, Blegdamsvej 17,
     DK-2100 Copenhagen {\O}, Denmark
    \label{NBI}}
\titlefoot{IPNP of MFF, Charles Univ., Areal MFF,
     V Holesovickach 2, CZ-180 00, Praha 8, Czech Republic
    \label{NC}}
\titlefoot{NIKHEF, Postbus 41882, NL-1009 DB
     Amsterdam, The Netherlands
    \label{NIKHEF}}
\titlefoot{National Technical University, Physics Department,
     Zografou Campus, GR-15773 Athens, Greece
    \label{NTU-ATHENS}}
\titlefoot{Physics Department, University of Oslo, Blindern,
     NO-1000 Oslo 3, Norway
    \label{OSLO}}
\titlefoot{Dpto. Fisica, Univ. Oviedo, Avda. Calvo Sotelo
     s/n, ES-33007 Oviedo, Spain
    \label{OVIEDO}}
\titlefoot{Department of Physics, University of Oxford,
     Keble Road, Oxford OX1 3RH, UK
    \label{OXFORD}}
\titlefoot{Dipartimento di Fisica, Universit\`a di Padova and
     INFN, Via Marzolo 8, IT-35131 Padua, Italy
    \label{PADOVA}}
\titlefoot{Rutherford Appleton Laboratory, Chilton, Didcot
     OX11 OQX, UK
    \label{RAL}}
\titlefoot{Dipartimento di Fisica, Universit\`a di Roma II and
     INFN, Tor Vergata, IT-00173 Rome, Italy
    \label{ROMA2}}
\titlefoot{Dipartimento di Fisica, Universit\`a di Roma III and
     INFN, Via della Vasca Navale 84, IT-00146 Rome, Italy
    \label{ROMA3}}
\titlefoot{DAPNIA/Service de Physique des Particules,
     CEA-Saclay, FR-91191 Gif-sur-Yvette Cedex, France
    \label{SACLAY}}
\titlefoot{Instituto de Fisica de Cantabria (CSIC-UC), Avda.
     los Castros s/n, ES-39006 Santander, Spain
    \label{SANTANDER}}
\titlefoot{Dipartimento di Fisica, Universit\`a degli Studi di Roma
     La Sapienza, Piazzale Aldo Moro 2, IT-00185 Rome, Italy
    \label{SAPIENZA}}
\titlefoot{Inst. for High Energy Physics, Serpukov
     P.O. Box 35, Protvino, (Moscow Region), Russian Federation
    \label{SERPUKHOV}}
\titlefoot{J. Stefan Institute, Jamova 39, SI-1000 Ljubljana, Slovenia
     and Laboratory for Astroparticle Physics,\\
     \indent~~Nova Gorica Polytechnic, Kostanjeviska 16a, SI-5000 Nova Gorica, Slovenia, \\
     \indent~~and Department of Physics, University of Ljubljana,
     SI-1000 Ljubljana, Slovenia
    \label{SLOVENIJA}}
\titlefoot{Fysikum, Stockholm University,
     Box 6730, SE-113 85 Stockholm, Sweden
    \label{STOCKHOLM}}
\titlefoot{Dipartimento di Fisica Sperimentale, Universit\`a di
     Torino and INFN, Via P. Giuria 1, IT-10125 Turin, Italy
    \label{TORINO}}
\titlefoot{Dipartimento di Fisica, Universit\`a di Trieste and
     INFN, Via A. Valerio 2, IT-34127 Trieste, Italy \\
     \indent~~and Istituto di Fisica, Universit\`a di Udine,
     IT-33100 Udine, Italy
    \label{TU}}
\titlefoot{Univ. Federal do Rio de Janeiro, C.P. 68528
     Cidade Univ., Ilha do Fund\~ao
     BR-21945-970 Rio de Janeiro, Brazil
    \label{UFRJ}}
\titlefoot{Department of Radiation Sciences, University of
     Uppsala, P.O. Box 535, SE-751 21 Uppsala, Sweden
    \label{UPPSALA}}
\titlefoot{IFIC, Valencia-CSIC, and D.F.A.M.N., U. de Valencia,
     Avda. Dr. Moliner 50, ES-46100 Burjassot (Valencia), Spain
    \label{VALENCIA}}
\titlefoot{Institut f\"ur Hochenergiephysik, \"Osterr. Akad.
     d. Wissensch., Nikolsdorfergasse 18, AT-1050 Vienna, Austria
    \label{VIENNA}}
\titlefoot{Inst. Nuclear Studies and University of Warsaw, Ul.
     Hoza 69, PL-00681 Warsaw, Poland
    \label{WARSZAWA}}
\titlefoot{Fachbereich Physik, University of Wuppertal, Postfach
     100 127, DE-42097 Wuppertal, Germany
    \label{WUPPERTAL}}
\addtolength{\textheight}{-10mm}
\addtolength{\footskip}{5mm}
\clearpage
\headsep 30.0pt
\end{titlepage}
%
\pagenumbering{arabic} 
\setcounter{footnote}{0} %
\large

\section{Introduction}

At LEP2, the Standard Model predicts that events with one or more photons and 
invisible particles are produced exclusively by the reaction 
$e^+e^-\rightarrow\nu\bar{\nu}\gamma(\gamma)$ which receives a contribution
from Z-exchange in the $s$-channel with single- or multi-photon emission 
from the initial state electrons and from the $t$-channel $W$ exchange, 
with the photon(s) radiated from the beam electrons or the exchanged $W$.

Beyond the Standard Model, contributions to the $\gamma~+~missing~energy$
final state could come from a new generation of neutrinos, 
from the radiative production of some 
new particle, stable or unstable, weakly interacting or
decaying into a photon.
Theories of supersymmetry (SUSY) predict the
existence of particles, such as the neutralino, which would produce
a final state with missing energy and a photon if the 
lightest neutralino decays into $\tilde{G}\gamma$ with an essentially massless
gravitino~\cite{GMSB,NLZ} and several results have been published on
the search for $\eeGGg$~\cite{LEP_results,delphi_sg133,delphi_161_172}.
If the gravitino is the only supersymmetric particle light enough to be produced,
the expected cross-section for $\eeGG$ can instead be used to set a lower limit
on the gravitino mass~\cite{gravitino}.

Also in the same SUSY theoretical framework, multi-photon final states 
with missing energy could be a signature for neutralino pair-production,
i.e. reactions of type $\eeGgGg$~\cite{GMSB,NLZ} and $\eeXgXg$~\cite{SUGRA}. 
In the case of long neutralino lifetimes the photons would not originate at the
beam interaction region and could have a large impact parameter.
For mean decay paths comparable to the detector scale, 
events with a single photon not pointing to the interaction region are expected.

In the study presented here, the single- and the multi-photon final states at LEP2
are used to explore the existence of possible new particles.
After a brief description of the detectors used in the analysis and
the selection criteria, a measurement of the number of neutrino families is made and
limits on non-Standard Model physics, such as high-dimensional 
gravitons~\cite{grav1,grav2}, compositeness~\cite{preons}
and supersymmetric particles, are presented.

This paper describes the analysis of single photon events collected
by DELPHI at centre-of-mass energies ($\sqrt{s}$) of 183~GeV and
189~GeV at LEP during 1997 and 1998. The integrated luminosities
at these energies were 51~pb$^{-1}$ and 158~pb$^{-1}$ respectively.
Single non-pointing photons and multi-photon events 
have also been studied, but in this case the analysis is restricted
to the data taken at $\sqrt{s}=$~189~GeV, since the results 
obtained at lower energies have already been published elsewhere~\cite{delphi_gg}.
The limits set on new phenomena take into account the lower energy data.

\section{The DELPHI detector}

The general criteria for the selection of events
are based mainly on the electromagnetic
calorimeters and the tracking system of the DELPHI detector~\cite{delphi_detec}.
All three major electromagnetic calorimeters in DELPHI,
the High density Projection Chamber (HPC),
the Forward ElectroMagnetic Calorimeter (FEMC) and the
Small angle TIle Calorimeter (STIC), have been used in the 
single-photon reconstruction. The barrel region is covered by the HPC, 
which is a gas sampling calorimeter able to sample a shower
nine times longitudinally. The FEMC is made up of an array of 
4532 lead glass blocks in each endcap. The energy resolution 
of this calorimeter is degraded by the material in front of it, 
which causes photon conversions and even preshowers.
The very forward luminosity monitor STIC~\cite{stic} consists of two
cylindrical lead-scintillator calorimeters read out by 
wavelength-shifting fibres. Two layers of 
scintillators mounted on the front of each STIC calorimeter together 
with a smaller ringshaped scintillator mounted directly on the beampipe,
provide $e - \gamma$ separation. 
The angular coverages of these calorimeters and the
energy resolutions are given in Table~\ref{calorimeters} and the
detailed characteristics and performances are described in~\cite{delphi_detec}.

Three different triggers are used in DELPHI to select single-photon events. 
The HPC trigger for purely neutral final states uses
a plane of scintillators inserted into one of the HPC
sampling gaps at a depth of around 4.5 radiation lengths.
A second level trigger decision is produced from the signals of 
analog electronics and is based on a coincidence pattern inside
the HPC module. 
The trigger efficiency has been measured with Compton and Bhabha
events. It is strongly dependent on the photon energy, $E_\gamma$, 
rising steeply up to $\sim$12~GeV,
with about $30\%$ efficiency at 4~GeV and above $80\%$ when $E_{\gamma}>30$~GeV.
It reaches a maximum of $87\%$ at $E_{\gamma} \simeq E_{\rm beam}$.
This efficiency does not include losses due to the cracks 
between modules of the HPC detector. 
The FEMC trigger requires an energy deposition of at least 2.5~GeV.
The efficiency increases with energy and is $\sim$97\% at 18~GeV. Correlated
noise in several adjacent channels causes fake triggers, but these
can be rejected offline with high efficiency by algorithms
that take into account the lead glass shower pattern. 
The STIC trigger requires an energy deposition of at least 15~GeV
and reaches maximum efficiency at 30~GeV. 
The trigger efficiency has been
measured with samples of photons from $e^+e^-\gamma$ and $q\bar{q}\gamma$
events. The efficiency varied between 74\% and 27\%
over the angular region used in the analysis.

In addition to the electromagnetic calorimeters, the DELPHI tracking system
was used to reject events in which charged particles are produced.
The main tracking devices are
the Time Projection Chamber (TPC) and the microVertex silicon Detector (VD) and its
extension into the forward region, the Very Forward Tracker (VFT). The silicon
trackers are also used for electron/photon separation by vetoing photon candidates
which can be associated with hits in these detectors. 

Finally, the Hadron CALorimeter (HCAL) and its cathode-read-out system
were used to reject cosmic rays and to provide photon/hadron separation, 
while the DELPHI Hermeticity Taggers were used to ensure complete detector 
hermeticity for additional neutral particles.

\begin{table}[hbt]
\begin{center}
\begin{tabular}{|c||c|c|c|c|}
\hline
  & Type & Angular coverage & $\sigma_{E}/E$ & X$_0$  \\
\hline
\hline
STIC: & Lead/scint. & $2^\circ <\theta<10^\circ$~,~$170^\circ <\theta<178^\circ$ & 
$0.0152\oplus(0.135/\sqrt{E})$  & 27 \\
\hline
FEMC: & Lead glass & $10^\circ <\theta<37^\circ$~,~$143^\circ <\theta<170^\circ$ & 
$0.03\oplus(0.12/\sqrt{E})\oplus(0.11/E)$ & 20 \\
\hline
HPC: & Lead/gas & $40^\circ <\theta < 140^\circ$ & 
$0.043\oplus(0.32/\sqrt{E})$ & 18 \\
\hline
\end{tabular}
\end{center}
\caption{Polar angle coverage, energy resolution (where $E$ is in GeV
and $\oplus$ denotes addition in quadrature)
and thickness (in radiation lengths) of the electromagnetic calorimeters
in DELPHI. }
\label{calorimeters}
\end{table}

\section{Event selection}

\subsection{Single-photon events}

The basic selection criteria of events were the same for the
three electromagnetic calorimeters:
no charged particle tracks detected and no electromagnetic showers apart 
from the tracks and showers caused by the single-photon candidate.  
However, the details of the
selection varied somewhat for the different electromagnetic calorimeters:
\begin{itemize}
\item
Events with a photon in the HPC were selected by requiring a 
shower having an energy above 6~GeV and a polar angle, $\theta$,
between $45^\circ$ and $135^\circ$ and no charged particle tracks.
The shower was required to satisfy conditions defining a 
good electromagnetic shape~\cite{delphi_sg133}.  
Background from radiative Bhabha events and Compton events
were rejected by requiring no other electromagnetic showers 
in the event unless they were in the HPC and within $20^\circ$
of the first one.
Cosmic rays were rejected mainly by the hadron calorimeter. If
there were two or more hadronic showers the event was discarded
and if only one HCAL shower was present, the event was rejected
if the shower was not consistent with being caused by punch-through of 
the electromagnetic shower. A constraint on the $\gamma$ direction 
was imposed, requiring that the line of flight from the mean interaction 
point and the shower direction measured in the calorimeter coincided within 
$15^\circ$. Also the requirement of no charged particles
removed cosmic ray background.
The photon identification efficiency depended on the criteria
applied to require a good electromagnetic shower. It was determined
on the basis of a Monte Carlo sample of 
events passed through the complete simulation of the DELPHI 
detector~\cite{delsim}.
The efficiency also depended on the photon energy and it
ranged from $\sim$45\% at 6~GeV to $\sim$71\% for 
$E_{\gamma}>15$~GeV. 

\item 
Events with at least one shower in the FEMC with an energy above 18~GeV and a
polar angle in the intervals $12^\circ<\theta<32^\circ$ or
$148^\circ<\theta<168^\circ$ were also selected.  Showers in the inner and outer radial
parts of the FEMC were discarded because of the 
large amount of material (about $2X_0$) in front of
the FEMC due to the STIC and the TPC detectors.  In order to separate electrons from
photons, the FEMC shower was extrapolated to the interaction point and the event
was rejected if hits in the silicon microvertex detectors (VD and VFT) could be
associated with the shower.

The material in front of the FEMC meant that about half of the photons preshowered
before reaching the calorimeter.  Most of the preshower was contained in a cone
of about 15$^\circ$ around the largest shower and the selection took this into
account by requiring no charged particle tracks, no other electromagnetic showers and no
hadronic showers outside a 15$^\circ$ cone.  If there were no charged particle tracks
inside the cone either, i.e., the photon had not preshowered, it was required
that only one FEMC shower was present in the event.  If, on the other hand,
charged particle tracks were present in the cone, more FEMC showers were allowed
and their momentum vectors were added to that of the largest shower.

The requirement of no electromagnetic showers outside the cone greatly
reduced the background of radiative Bhabha and Compton events by rejecting
events that had one or both electrons in the acceptance of the experiment.
Events due to cosmic rays were rejected by the requirement of no
hadronic showers outside the cone. Inside the cone, hadronic energy was
allowed only in the first layer of the HCAL.

Most reconstruction and event selection efficiencies in the analysis were
taken into account by using Monte Carlo samples passed through the
extensive detector simulation package of DELPHI~\cite{delsim}. 
Some efficiencies, however, were determined from data. In particular,
the requirements of no electromagnetic or hadronic showers and no charged
particles were studied. A sample of events triggered at random and a sample
of back-to-back Bhabha events with the electrons in the STIC were used for this purpose. 
It was found that noise and machine background caused showers and tracks which would
veto about 14\% of the good single-photon events. 

\item
Single photons in the STIC were preselected by requiring 
one shower with an energy of at least 27~GeV in one of the two
STIC calorimeters and with $3.8^\circ<\theta<8^\circ$ or
$172^\circ<\theta<176.2^\circ$ and no other electromagnetic showers, no hadronic
showers and no charged particles in the event.
It was furthermore required that all single-photon candidates had satisfied
the STIC single-photon trigger and that there was no signal in at least one
of the two scintillator planes in front of the shower. 
A requirement of no signal in the small scintillators mounted on 
the beampipe made it possible to reject some of the radiative
$ee\gamma$ background. In spite of the scintillator requirements,
the huge background of off-energy electrons made it necessary
to introduce a $\theta$-dependent energy cut in such a way that
$x_{\gamma} > (9.2^\circ -\theta)/9^\circ$ for $\theta<6.5^\circ$
where $x_{\gamma}=E_{\gamma}/E_{beam}$.
 
The trigger efficiency in the STIC acceptance was discussed in Section~2.
The offline photon identification and reconstruction resulted in an additional 
loss of 5\% of the photons.
The selection of events with no shower in the STIC and no tracks 
implied similar losses to those found in the FEMC analysis and were estimated
with the same methods. 

\end{itemize}

\subsection{Non-pointing single-photon events}

The fine granularity of the HPC calorimeter provided a 
precise reconstruction of the
axis direction in electromagnetic showers. This feature
was used to select events with a single photon whose
flight direction did not point to the beam interaction region.
Events with a single non-pointing photon are expected 
when two neutral particles with large mean decay paths ($> 4$ m) 
are produced which subsequently decay into a photon and an invisible particle.

Events of this kind were searched for by requiring one photon
in the HPC calorimeter with $E_\gamma>10$~GeV and impact parameter
exceeding 40 cm. Cosmic ray events, which represent the main
experimental background, were largely reduced by vetoing on
isolated hits or tracks in the Hermeticity Taggers and signals from
the cathode-read-out system of the hadron calorimeter.
More details on the precise event selection can be found 
in~\cite{delphi_gg}, where the analysis of the data samples collected
at centre-of-mass energies up to 183~GeV is described.
The same analysis has been applied to the data sample taken
at 189~GeV.

\subsection{Multi-photon events}

A study of final states with at least two photons and
missing energy at $\sqrt{s}=189$~GeV has also been made.

As for non-pointing single photons,
the physics motivations and the selection criteria 
have been discussed in detail in the published paper~\cite{delphi_gg}
dedicated to the analysis of the data taken at centre-of-mass
energies up to 183~GeV. Here only a brief update of the results is
given using the 189~GeV data and the same analysis method.

The selection of multi-photon final states was, as in the 183~GeV analysis,
based on a two-step procedure:

\begin{itemize}
\item In a first step all events with missing transverse energy and
at least two photons, each with $x_\gamma>0.05$ (where
$x_\gamma =E_\gamma / E_{beam}$), were preselected.
Very loose cuts on the polar angle of the photon and
acoplanarity were adopted for the selection of this sample,
which was used to monitor the modelling of the 
$\ee \to \nu \nu \gamma \gamma (\gamma)$ process by the KORALZ 4.02
generator~\cite{koralz}. 
\item In a second step these criteria were tightened in order to improve the 
experimental sensitivity for possible signals of supersymmetry, such as
the $\eeGgGg$ or $\eeXgXg$ processes.
This was achieved by imposing more stringent requirements
on the photon polar angles as well as on the event missing mass and
transverse momentum.
\end{itemize}

More details on the event selection can be found in~\cite{delphi_gg}.

\section{Real and simulated data samples}

Apart from the $\eenng (\gamma)$ process, single-photon 
events can be faked by the QED reaction
$e^+e^- \rightarrow e^+e^-\gamma $
if the two electrons escape undetected along the beampipe or
if the electrons are in the detector acceptance but are not
detected by the experiment.

This process has a very high cross-section,
decreasing rapidly when the energy ($E_{\gamma}$) and the
polar angle ($\theta_{\gamma}$) of the photon increase. 
The behaviour of this QED background 
together with the rapidly varying efficiencies at low energies
are the reasons why different energy cuts had to be applied for
photons in the three calorimeters. In the final analysis it was required 
that $x_{\gamma}>0.06$ (HPC) and $x_{\gamma}>0.2$ (FEMC). In the STIC analysis,
the requirement was $x_{\gamma}>0.3$ for $6.5^\circ<\theta<8.0^\circ$
and $x_{\gamma}>(9.2^\circ -\theta)/9^\circ$ for $3.8^\circ < \theta<6.5^\circ$.

The critical parameter in the rejection of the $e^+e^-\gamma$ background is 
the polar angle
at which the electrons start being seen in
the STIC detector. This detector reconstructs electrons down to
$\theta =2.2^\circ$ and in addition, the scintillator counters mounted on the beampipe can 
be used to reject events with electrons down to $1.8^\circ$. Simulations 
have shown that even at lower angles (down to $0.97^\circ$) a large fraction of the electrons 
are detectable because they interact with a tungsten shield mounted inside
the beampipe and leak enough energy into the STIC to make it possible
to reject the events. 

The remaining background from the $e^+e^-\gamma$ process 
was calculated with a 
Monte Carlo program~\cite{TEEGG} and two different event topologies 
were observed. 
Either both electrons were below the STIC acceptance or one of the electrons
was in the DELPHI acceptance where it was wrongly identified as a photon, and the
photon was lost in the cracks between the electromagnetic calorimeters.
The first topology gives background at low photon energy while the
second one produces fake photon events
at high energy. In the HPC acceptance an analytical
calculation~\cite{eeg} was also used to confirm
that the $e^+e^-\gamma$ background was negligible.

In the STIC analysis, an additional background is the single electrons produced by 
interactions between the beam
particles and residual gas molecules in the LEP beampipe. In these
$e\rightarrow e\gamma$ events the photons are always lost in the beampipe
while the off-energy electrons are bent into the STIC acceptance by
the low-beta quadrupoles close to DELPHI. The rate of this background is so large
that it was not possible to provide a $\gamma-e$ separation powerful
enough to eliminate this background completely. A simulation
has been made of off-energy electron production~\cite{off},
but it could not be used in the analysis since the vacuum pressure 
around the LEP ring was not known to the required precision.
Instead, a background sample was collected with a trigger similar to the 
photon trigger except that it did not use the scintillators for photon-electron
separation. 
After applying all the cuts used in the single photon analysis, except the scintillator
requirements, this background sample was used to estimate the remaining off-energy
electron background.

The contribution from other processes such as $\gamma\gamma$ collisions,
$e^+e^- \rightarrow \gamma \gamma \gamma$, cosmic ray events, 
$e^+e^- \rightarrow \mu^+ \mu^- \gamma$ and 
$e^+e^- \rightarrow \tau^+ \tau^- \gamma $
has also been calculated.

The $\nu\bar{\nu}\gamma(\gamma)$ process was simulated by both the KORALZ~\cite{koralz}
and the NUNUGPV~\cite{nunugpv} program with very similar results
(the numbers of expected events in
the HPC region at 189~GeV were estimated to be 156.8 and 157.7 with the two 
programs respectively).

A detailed discussion on the backgrounds for the non-pointing single-photon
events and for the multi-photon events is contained in~\cite{delphi_gg}. 

\begin{table}[hbt]
\begin{center}
\begin{tabular}{|c|c|c|c|c|c|c|}
\cline{2-7}
\multicolumn{1}{c||}{  }    & 
\multicolumn{2}{|c|}{ HPC }  & 
\multicolumn{2}{|c|}{ FEMC}  & 
\multicolumn{2}{|c|}{ STIC } \\
\hline
\multicolumn{1}{|c||}{$\theta_{\gamma}$:} & 
\multicolumn{2}{|c|}{$45^\circ -135^\circ$}  & 
\multicolumn{2}{|c|}{$12^\circ -32^\circ$ , $148^\circ -168^\circ$} & 
\multicolumn{2}{|c|}{$3.8^\circ -8.0^\circ$ , $172^\circ -176.2^\circ$} \\
\multicolumn{1}{|c||}{$x_{\gamma}$:} & 
\multicolumn{2}{|c|}{$>$ 0.06}  & 
\multicolumn{2}{|c|}{0.2 - 0.9}  & 
\multicolumn{2}{|c|}{0.3 - 0.9} \\
\hline\hline
\multicolumn{1}{|c||}{ $\sqrt{s}$: } & 
\multicolumn{1}{|c|}{ 182.7~GeV }      & \multicolumn{1}{|c|}{ 188.7~GeV } &
\multicolumn{1}{|c|}{ 182.7~GeV }      & \multicolumn{1}{|c|}{ 188.7~GeV } &
\multicolumn{1}{|c|}{ 182.7~GeV }      & \multicolumn{1}{|c|}{ 188.7~GeV } \\
\hline
\multicolumn{1}{|c||}{ Luminosity: } & 
\multicolumn{1}{|c|}{ 50.2 $pb^{-1}$ }      & \multicolumn{1}{|c|}{ 154.7 $pb^{-1}$ } &
\multicolumn{1}{|c|}{ 49.2 $pb^{-1}$ }      & \multicolumn{1}{|c|}{ 157.7 $pb^{-1}$ } &
\multicolumn{1}{|c|}{ 51.4 $pb^{-1}$ }      & \multicolumn{1}{|c|}{ 157.3 $pb^{-1}$ } \\
\hline\hline
\multicolumn{1}{|c||}{ $N_{observed}$: } & 
\multicolumn{1}{|c|}{ 54 }      & \multicolumn{1}{|c|}{146 } &
\multicolumn{1}{|c|}{ 65 }      & \multicolumn{1}{|c|}{155 } &
\multicolumn{1}{|c|}{ 32 }      & \multicolumn{1}{|c|}{ 94 } \\
\hline
\multicolumn{1}{|c||}{ $N_{background}$: } & 
\multicolumn{1}{|c|}{ 0.08 }     & \multicolumn{1}{|c|}{ 0.3 } &
\multicolumn{1}{|c|}{ 3.5 }      & \multicolumn{1}{|c|}{ 6.0 } &
\multicolumn{1}{|c|}{ 3.6 }      & \multicolumn{1}{|c|}{ 6.5 } \\
\hline
\multicolumn{1}{|c||}{ $N_{e^+e^-\rightarrow\nu\bar{\nu}\gamma}$: } & 
\multicolumn{1}{|c|}{ 59.5$\pm$1.6 }      & \multicolumn{1}{|c|}{ 156.8$\pm$4.3 } &
\multicolumn{1}{|c|}{ 55.0$\pm$1.2 }      & \multicolumn{1}{|c|}{ 153.4$\pm$1.9 } &
\multicolumn{1}{|c|}{ 32.4$\pm$0.7 }      & \multicolumn{1}{|c|}{  91.4$\pm$0.9 } \\
\hline\hline
\multicolumn{1}{|c||}{ $\sigma_{meas}$~(pb) } & 
\multicolumn{1}{|c|}{ 1.85$\pm$0.25 }      & \multicolumn{1}{|c|}{ 1.80$\pm$0.15 } &
\multicolumn{1}{|c|}{ 2.33$\pm$0.31 }      & \multicolumn{1}{|c|}{ 1.89$\pm$0.16 } &
\multicolumn{1}{|c|}{ 1.27$\pm$0.25 }      & \multicolumn{1}{|c|}{ 1.41$\pm$0.15 } \\
\hline
\multicolumn{1}{|c||}{ $\sigma_{\nu\bar{\nu}\gamma(\gamma)}$~(pb) } & 
\multicolumn{1}{|c|}{ 2.04 }      & \multicolumn{1}{|c|}{ 1.97 } &
\multicolumn{1}{|c|}{ 2.08 }      & \multicolumn{1}{|c|}{ 1.94 } &
\multicolumn{1}{|c|}{ 1.50 }      & \multicolumn{1}{|c|}{ 1.42 } \\
\hline
\multicolumn{1}{|c||}{ $N_{\nu}$ } & 
\multicolumn{1}{|c|}{ 2.63$\pm$0.49 }    & \multicolumn{1}{|c|}{ 2.65$\pm$0.31 } &
\multicolumn{1}{|c|}{ 3.42$\pm$0.51 }    & \multicolumn{1}{|c|}{ 2.91$\pm$0.28 } &
\multicolumn{1}{|c|}{ 2.49$\pm$0.57 }    & \multicolumn{1}{|c|}{ 2.98$\pm$0.37 } \\
\hline
\end{tabular}
\end{center} 
\caption{Number of selected and expected single photon events,
measured and calculated cross-section for 
$e^+e^- \rightarrow \nu \bar{\nu}\gamma(\gamma)$
(KORALZ with three neutrino generations) and the number of neutrino generations
calculated from the cross-sections. The errors are statistical only.
$x_\gamma$ is $E_\gamma / E_{beam}$.}
\label{cross} 
\end{table}

\begin{figure}[htb]
\centerline{\epsfig
{file=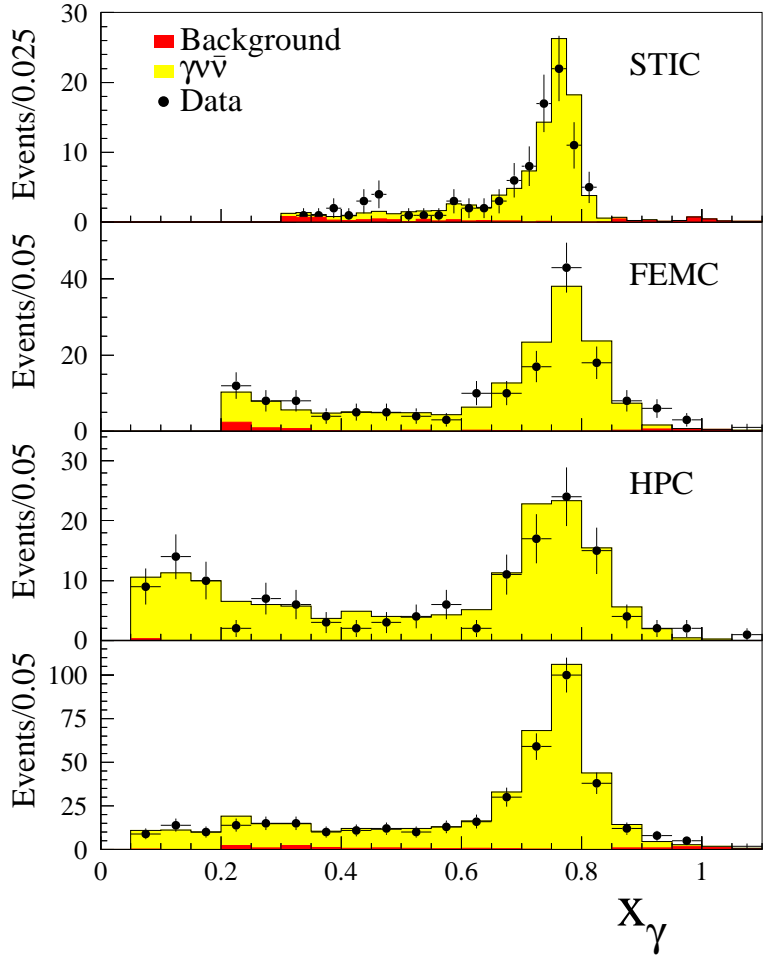,width=10cm}}
\caption[]{
$x_{\gamma}$ of selected single photons at 189~GeV in 
the three calorimeters STIC, FEMC and HPC. The bottom
plot shows the combined spectrum.
The light shaded area is the expected distribution from
$e^+e^- \rightarrow \nu \bar{\nu}\gamma$ and the
dark shaded area is the total background from other
sources.}
\label{somma_xgam} 
\end{figure}

\begin{table}[hbt]
\begin{center}
\begin{tabular}{|c||c|c|c|c|c|c|}
\cline{2-7}
\multicolumn{1}{c||}{        } & \multicolumn{2}{|c|}{ HPC } & \multicolumn{2}{|c|}{ FEMC} & \multicolumn{2}{|c|}{ STIC } \\
\hline
\multicolumn{1}{|c||}{ Source } & 
\multicolumn{1}{|c|}{ Variation }  & \multicolumn{1}{|c|}{ $\Delta\sigma$ } &
\multicolumn{1}{|c|}{ Variation }  & \multicolumn{1}{|c|}{ $\Delta\sigma$ } &
\multicolumn{1}{|c|}{ Variation }  & \multicolumn{1}{|c|}{ $\Delta\sigma$ } \\
\hline
\multicolumn{1}{|l||}{ Luminosity } & 
\multicolumn{1}{|c|}{ $\pm$0.6\% }  & \multicolumn{1}{|c|}{ $\pm$0.6\% } &
\multicolumn{1}{|c|}{ $\pm$0.6\% }  & \multicolumn{1}{|c|}{ $\pm$0.6\% } &
\multicolumn{1}{|c|}{ $\pm$0.6\% }  & \multicolumn{1}{|c|}{ $\pm$0.6\% } \\
\multicolumn{1}{|l||}{ Trigger efficiency } & 
\multicolumn{1}{|c|}{ $\pm$5\% }  & \multicolumn{1}{|c|}{ $\pm$5\% } &
\multicolumn{1}{|c|}{ $\pm$2\% }  & \multicolumn{1}{|c|}{ $\pm$2\% } &
\multicolumn{1}{|c|}{ $\pm$6\% }  & \multicolumn{1}{|c|}{ $\pm$6\% } \\
\multicolumn{1}{|l||}{ Identification efficiency } & 
\multicolumn{1}{|c|}{ $\pm$5\% }  & \multicolumn{1}{|c|}{ $\pm$5\% } &
\multicolumn{1}{|c|}{ $\pm$6\% }  & \multicolumn{1}{|c|}{ $\pm$6\% } &
\multicolumn{1}{|c|}{ $\pm$5\% }  & \multicolumn{1}{|c|}{ $\pm$5\% } \\
\multicolumn{1}{|l||}{ Calorimeter energy scale } & 
\multicolumn{1}{|c|}{ $\pm$5\% }  & \multicolumn{1}{|c|}{ $\pm$4\% } &
\multicolumn{1}{|c|}{ $\pm$4\% }  & \multicolumn{1}{|c|}{ $\pm$4\% } &
\multicolumn{1}{|c|}{ $\pm$0.5\% }  & \multicolumn{1}{|c|}{ $\pm$1\% } \\
\multicolumn{1}{|l||}{ Background } & 
\multicolumn{1}{|c|}{ $\pm$57\% }  & \multicolumn{1}{|c|}{ $\pm$0.1\% } &
\multicolumn{1}{|c|}{ $\pm$55\% }  & \multicolumn{1}{|c|}{ $\pm$2\% } &
\multicolumn{1}{|c|}{ $\pm$62\% }  & \multicolumn{1}{|c|}{ $\pm$5\% } \\
\hline
\multicolumn{1}{|l||}{ Total } & 
\multicolumn{1}{|c|}{ }  & \multicolumn{1}{|c|}{ $\pm$8\% } &
\multicolumn{1}{|c|}{ }  & \multicolumn{1}{|c|}{ $\pm$8\% } &
\multicolumn{1}{|c|}{ }  & \multicolumn{1}{|c|}{ $\pm$9\% } \\
\hline
\end{tabular}
\end{center} 
\caption{Contributions to systematic error.
The total systematic error is the quadratic sum
of the individual errors.} 
\label{syst} 
\end{table}

\section{Comparison with the Standard Model expectations}

\subsection{Single-photon cross-section}

The final numbers of expected and observed single-photon events are given in
Table~\ref{cross} and the $x_{\gamma}$ spectrum of the selected events at 189~GeV is
shown in Figure~\ref{somma_xgam} together with the expected
background and the $\nu\bar{\nu}\gamma$ contribution.
The single-photon event selection was such that
events with more than one photon could survive
if the other photons were at low angle ($\theta_{\gamma} < 2.2^\circ$), 
low energy ($E_{\gamma} <$ 0.8 GeV)
or within 3$^\circ$, 15$^\circ$ and 20$^\circ$ from
the highest energy photon in the
STIC, FEMC and HPC respectively.
In total, 546 single-photon events were observed at 189~GeV and 183~GeV in
the three calorimeters, with 570 events expected from known sources. 

\begin{figure}[htb]
\centerline{\epsfig
{file=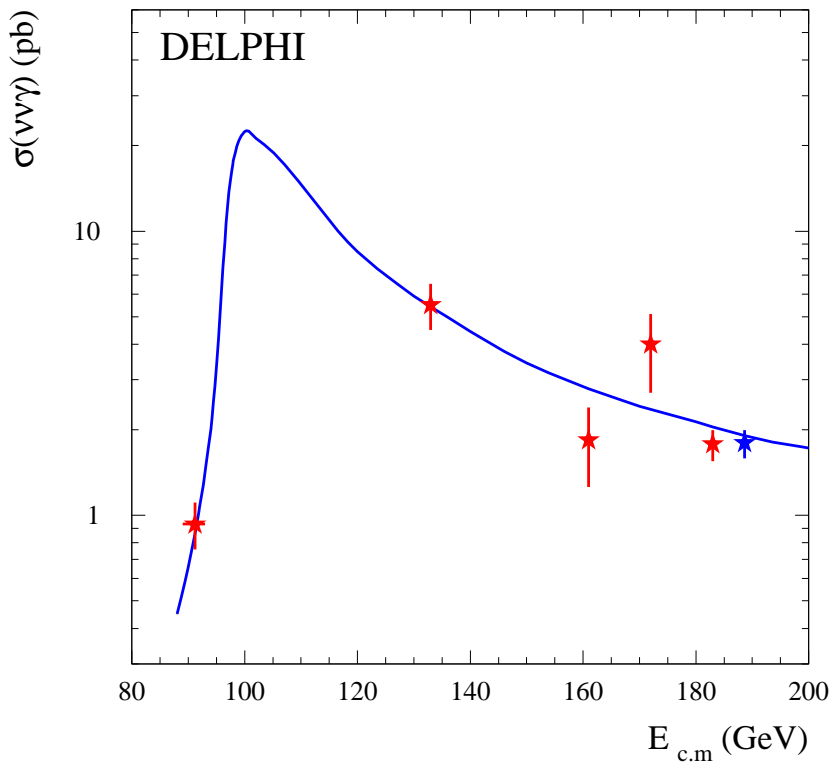,width=12.cm}}
\caption[]{
The measured cross-sections in the HPC for E$_\gamma>$6~GeV
at different $\sqrt{s}$
compared to the expected $\sigma(\nu \bar\nu \gamma)$ 
(for three neutrino generations).}
\label{sigma6}
\end{figure}

The measured cross-sections calculated from the single-photon events after
correcting for background and efficiencies are given in Table~\ref{cross}. 
The previously mentioned Monte Carlo programs were used to 
calculate the expected values of the cross-section of the process 
$e^+e^- \rightarrow \nu \bar{\nu}\gamma(\gamma)$ inside the acceptance of each of the 
three detectors used in the analysis.
Figure~\ref{sigma6} shows the expected behaviour of the cross-section,
calculated with NUNUGPV for three neutrino generations, compared with the values measured 
with the HPC detector at different LEP energies. 
The contributions from various sources to the systematic error in the 
cross-section measurement are given in Table \ref{syst}.
The dominant uncertainty comes from the estimation of trigger and detection efficiencies.
The calculation of the expected cross-section has a theoretical uncertainty 
which is approaching 1\% with the latest versions of
NUNUGPV~\cite{nunugpv} and KORALZ~\cite{koralz} and this error is thus insignificant 
compared with the experimental systematic errors.

A measurement of the cross-section of the process $e^+e^- \rightarrow \nu \bar{\nu}\gamma$ 
determines the number of light neutrino generations, $N_{\nu}$. DELPHI has
previously reported a value of $N_{\nu}=2.89\pm0.32$ from LEP1 
single photon data~\cite{delphi_sg92}. 
The LEP2 cross-section measurements have now been compared with the expected
cross sections for 2, 3 and 4 neutrino generations, calculated with KORALZ,
and the number of neutrino generations has been deduced (Table~\ref{cross}).
Averaging the three independent measurements from the three different
calorimeters at 183~GeV and 189~GeV, 
the number of light neutrino generations becomes:
$$N_{\nu}=2.84\pm0.15(stat)\pm0.14(syst)$$

\subsection{Non-pointing single-photon events and multi-photon events} 

The numbers of events with a single non-pointing photon or with 
multi-photon final states found in the data sample at 189~GeV 
are compared to Standard Model expectations in Table~\ref{samples}.

\begin{figure}[hbt]
\begin{center}
 \mbox{\epsfxsize 7.8cm \epsfbox{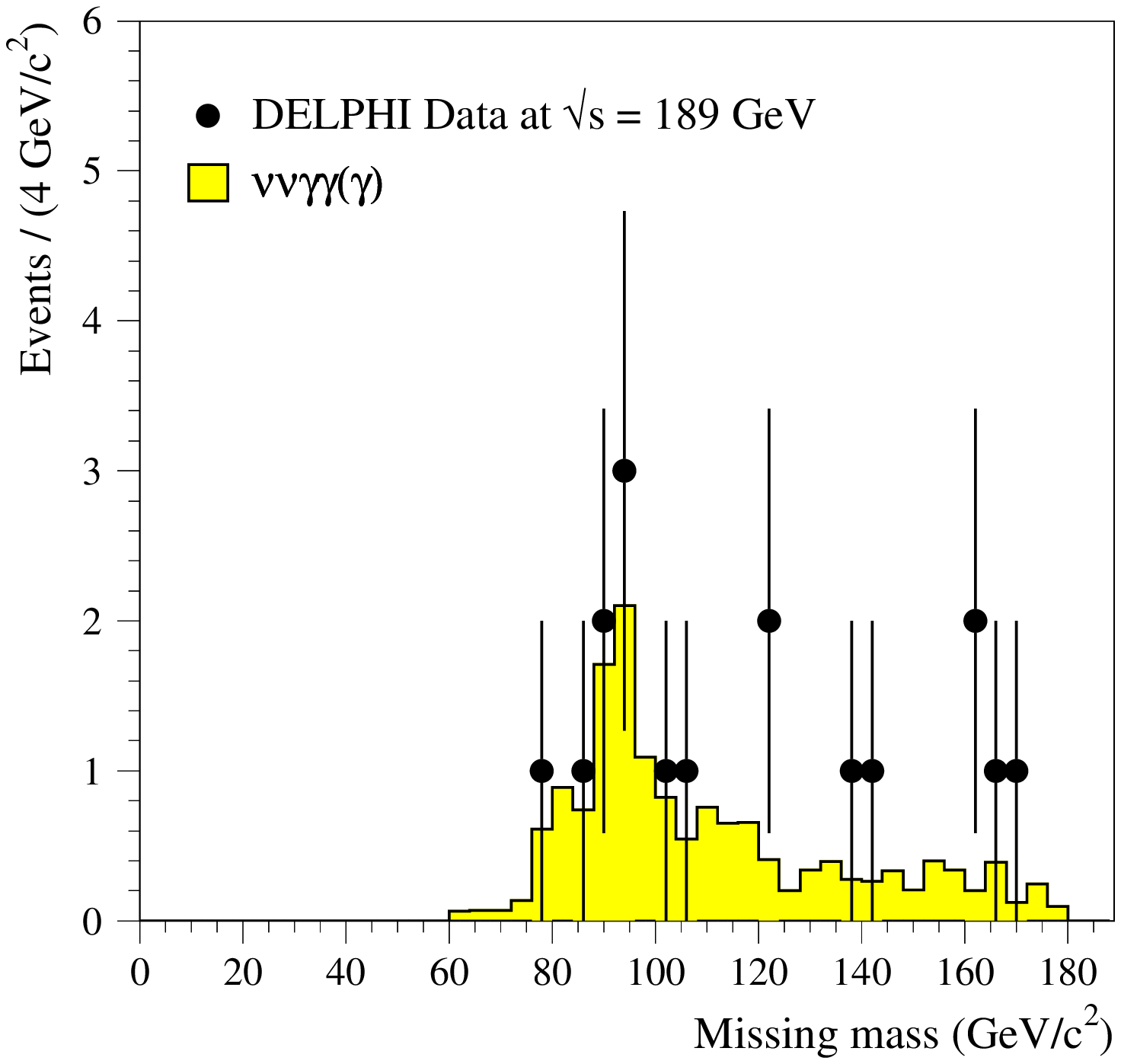}}
 \mbox{\epsfxsize 7.8cm \epsfbox{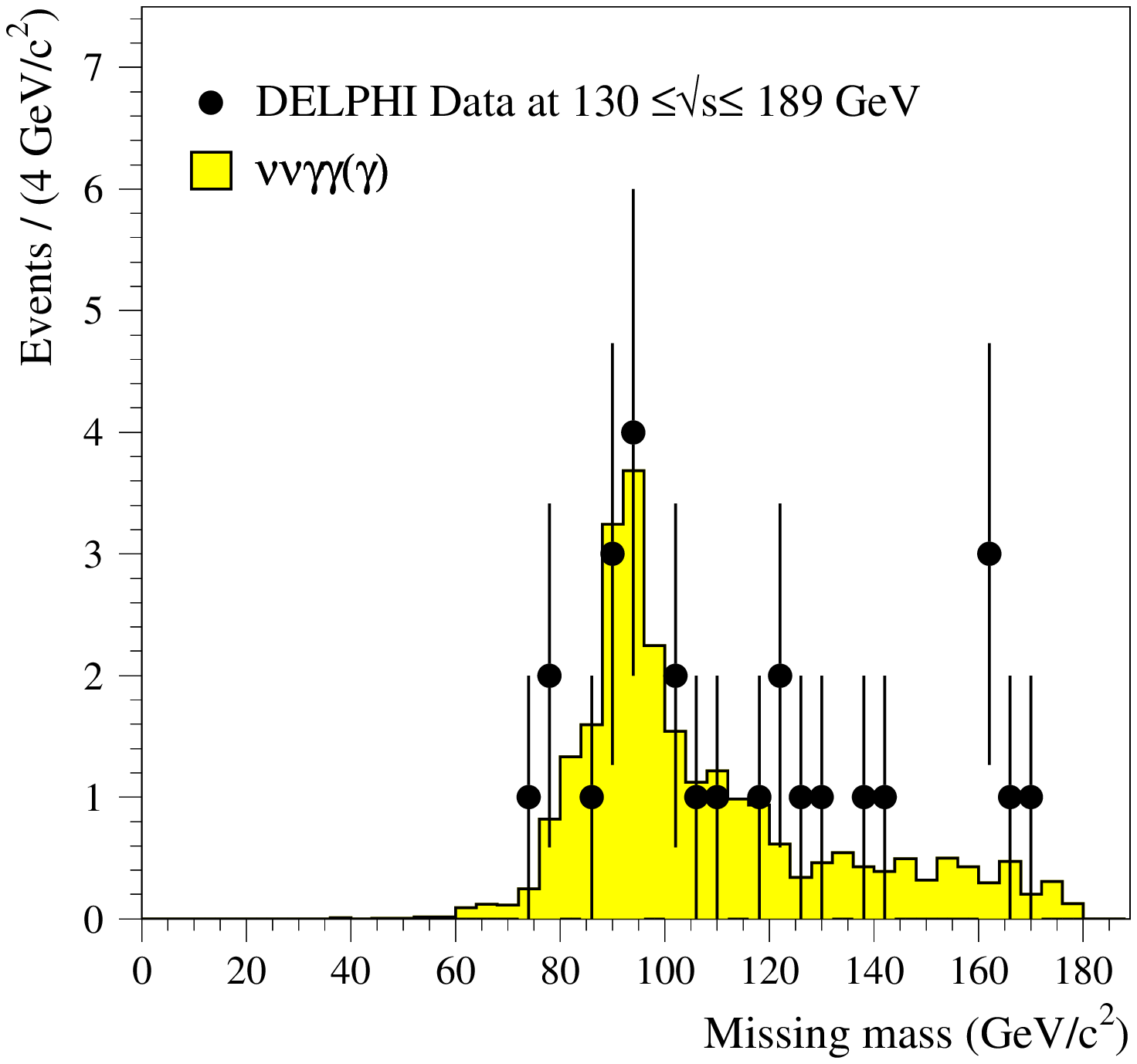}}
\end{center}
\caption[]{Missing mass distribution observed after multi-photon 
preselection in the 189~GeV sample
(left) and the combined 130-189~GeV sample (right).}
\label{preselection}
\end{figure}

The missing mass spectra for
the preselected multi-photon events and the expected contribution from
$e^+e^-\rightarrow\nu\overline{\nu}\gamma\gamma(\gamma)$ as
simulated with KORALZ are shown in 
Figure~\ref{preselection}. The measured missing mass distribution 
is in good agreement with the simulation.

No excess over Standard Model expectations was found in 
any of the data samples collected at $\sqrt{s}=189$~GeV. Hence
these data were combined with lower energy data to extract 
limits on new physics. 

\begin{table}[hbt]
\begin{center}
\begin{tabular}{|c|c|c|c|c|}
\cline{2-5}
\multicolumn{1}{c|}{ }    & 
\multicolumn{2}{|c||}{ 189~GeV }  & 
\multicolumn{2}{|c|}{ 130-189~GeV }  \\
\cline{2-5}
\multicolumn{1}{c|}{  } & 
\multicolumn{1}{|c|}{ Observed }      & \multicolumn{1}{|c||}{ Expected } &
\multicolumn{1}{|c|}{ Observed }      & \multicolumn{1}{|c|}{ Expected } \\
\hline
\multicolumn{1}{|c|}{ Preselected multi-photon events } & 
\multicolumn{1}{|c|}{ 17 }      & \multicolumn{1}{|c||}{ 15.1$\pm$0.9 } &
\multicolumn{1}{|c|}{ 27 }      & \multicolumn{1}{|c|}{ 25.3$\pm$1.0 } \\
\hline
\multicolumn{1}{|c|}{ $\eeGgGg$ selection } & 
\multicolumn{1}{|c|}{ 5 }      & \multicolumn{1}{|c||}{ 4.4$\pm$0.5 } &
\multicolumn{1}{|c|}{ 7 }      & \multicolumn{1}{|c|}{ 7.1$\pm$0.5 } \\
\hline
\multicolumn{1}{|c|}{ $\eeXgXg$ selection } & 
\multicolumn{1}{|c|}{ 8 }      & \multicolumn{1}{|c||}{ 5.2$\pm$0.5 } &
\multicolumn{1}{|c|}{ 12 }     & \multicolumn{1}{|c|}{ 8.6$\pm$0.6 } \\
\hline
\multicolumn{1}{|c|}{ Non-pointing single-photon events } & 
\multicolumn{1}{|c|}{ 4 }      & \multicolumn{1}{|c||}{ 5.0$\pm$0.6 } &
\multicolumn{1}{|c|}{ 6 }      & \multicolumn{1}{|c|}{ 7.6$\pm$0.9 } \\
\hline
\end{tabular}  
\end{center} 
\caption[]{The number of observed and expected events from Standard Model sources 
in four selected data samples.}
\label{samples} 
\end{table}

\begin{figure}[hbt]
\begin{center}
 \mbox{\epsfxsize 7.8cm \epsfbox{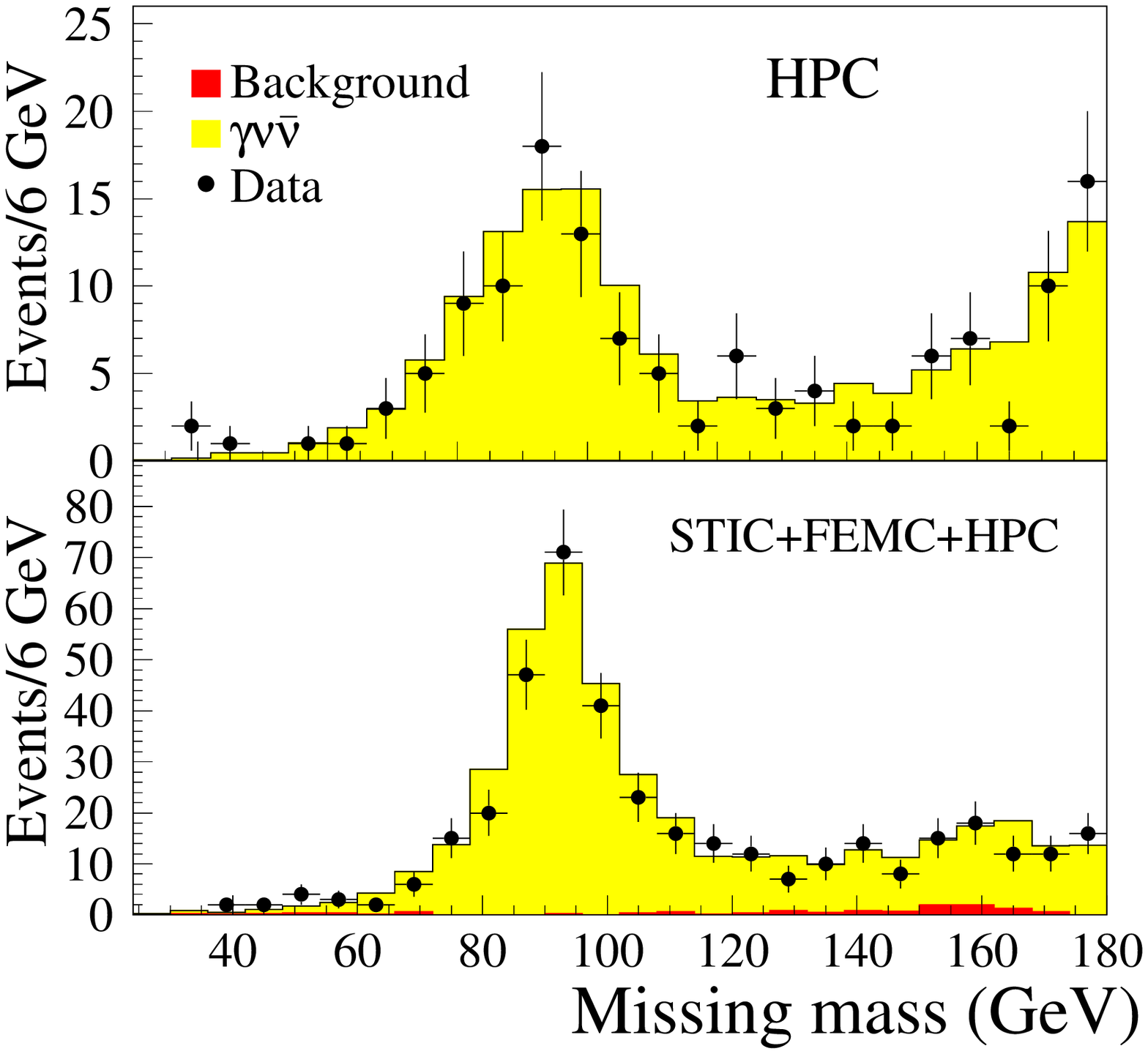}}
 \mbox{\epsfxsize 7.8cm \epsfbox{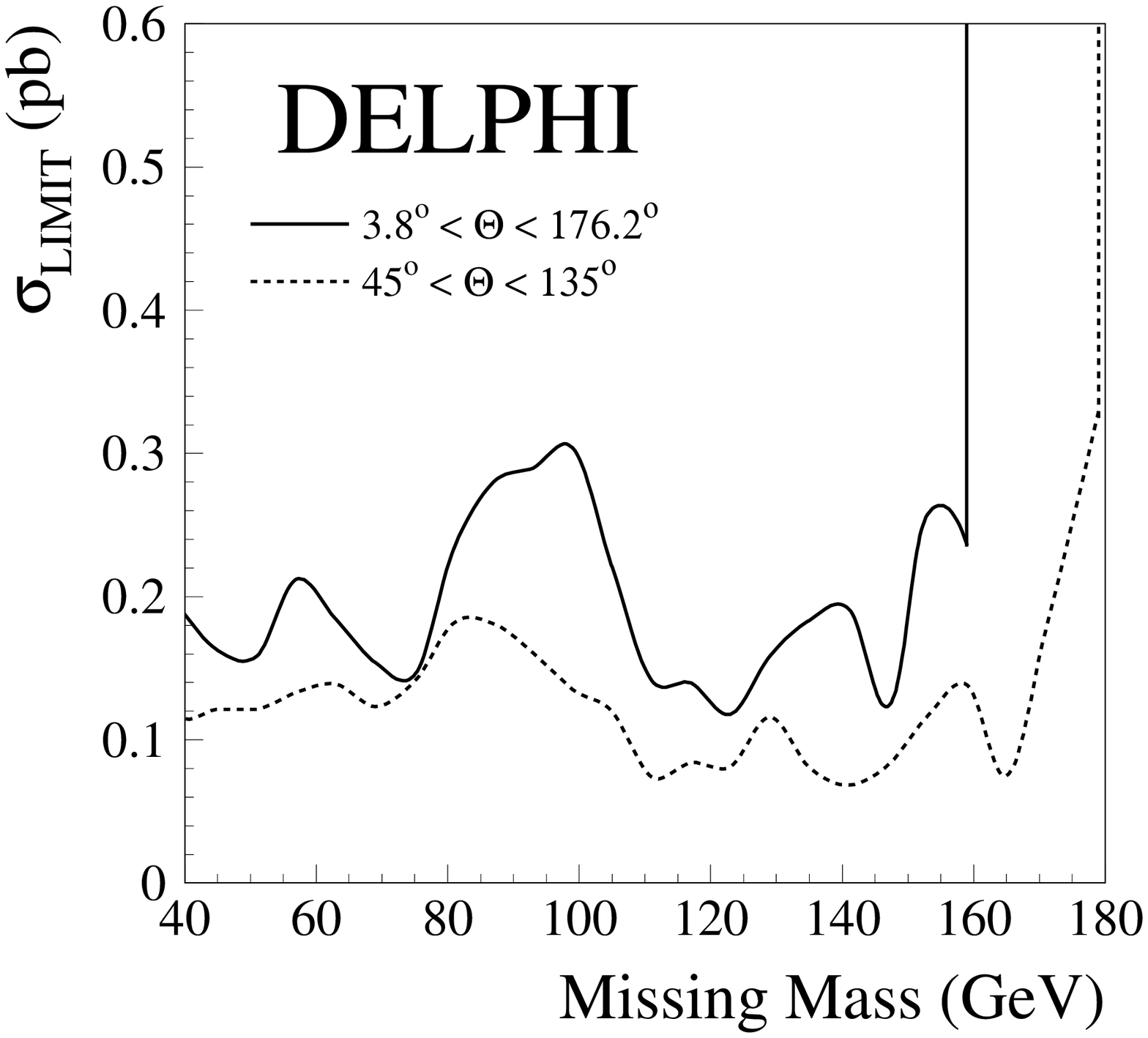}}
\end{center}
\caption[]{Left: The distributions of the missing mass 
for the events at 189~GeV in the HPC and in all three calorimeters.
The light shaded area is the expected distribution from
$e^+e^- \rightarrow \nu \bar{\nu}\gamma$ and the
dark shaded area is the total background from other
sources. Right: upper limit at 95$\%$~C.L. (within the solid angles described) 
for the production of a new unknown stable neutral object .
}
\label{mrecl}
\end{figure}

\section{Limits on new phenomena}

\subsection{Limits on the production of an unknown neutral state}

In many previous analyses~\cite{delphi_sg133,delphi_sg92,opal_sg92}
the observed single-photon candidates have been used to set
a limit on the probability of the existence of a
new particle, X, produced in association with a photon
and being stable or decaying into invisible particles.
The limit is calculated from the missing mass distribution 
(Figure~\ref{mrecl})
of the 395 single photon events at 189~GeV in the $\gamma$ angular region 
$3.8^\circ <\theta < 176.2^\circ$,
while taking into account the expected contributions from the Standard Model.
The limit is valid when the intrinsic width of the X particle is negligible
compared with the detector resolution
(the missing mass resolution varies between 10~GeV/c$^2$ at the $Z^0$ peak 
to 1~GeV/c$^2$ at high masses).
The upper limit at the 95\% confidence
level of the cross-section for 
$e^+e^- \rightarrow \gamma +$X 
is given in Figure~\ref{mrecl} for photons in the HPC region and
in all three calorimeters combined. In the latter case an assumption of an ISR-like
photon angular distribution has been made to correct for the regions between
the calorimeters.

\begin{figure}[hbt]
\centerline{\epsfig {file=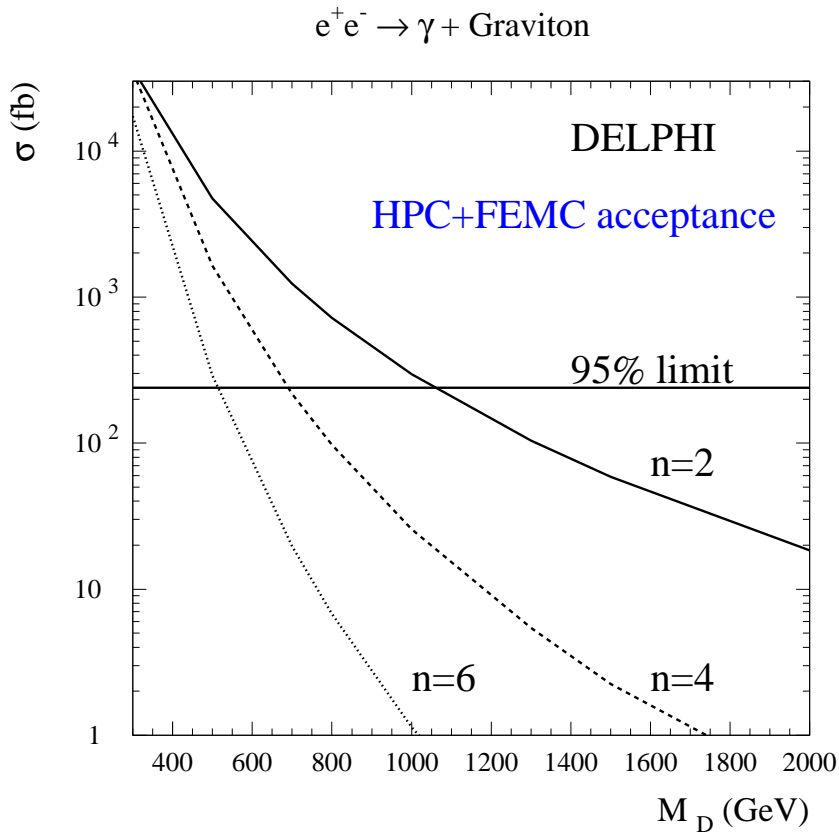,width=12cm}}
\caption[]{The cross-section limit at 95\% C.L. 
for $e^+e^-\rightarrow\gamma G$ production
and the expected cross-section for 2, 4 and 6 extra dimensions.} 
\label{grav} 
\end{figure}

\subsection{Limits on the production of gravitons}

It has been suggested recently~\cite{grav1,grav2} 
that gravitational interactions could be
unified with gauge interactions already at the weak scale if there are
extra compact dimensions of space in which only gravity can propagate.
The observed weakness of gravitation compared to other forces would be
related to the size of the compactified extra dimensions. A fundamental
mass scale $M_{D}$ is introduced, which is related to the gravitational constant 
$G_{N}$ and to the size or radius $R$ of the compactified space
(assumed to be a torus) by 
$$ M_{D}^{n+2}R^n = (8\pi G_N)^{-1}$$
where $n$ is the number of dimensions in addition to the usual
4 dimensional space. With one extra dimension and a fundamental
scale of 0.5-1~TeV, the size of this dimension becomes
$10^{12}-10^{13}$~m which is excluded by macroscopic measurements.
However, already with two extra dimensions, $R$ is in the range 0.5-1.9~mm
and with $n$=6 the size of the dimensions becomes 0.3-0.7 \AA . 
In this case the modification of the gravitational force
would not have been observed in previous gravitational measurements. 

The consequence of this model is that at LEP gravity could manifest 
itself by the production of gravitons ($G$), which themselves 
would be undetectable by the experiments. Instead 
single photons from the $e^+e^-\rightarrow\gamma G$ reaction 
are observable. 
The differential cross-section for this process has been 
calculated~\cite{grav2}. 
Most of the signal
is expected at low photon energy and, since 
$\sigma\sim s^{n/2}/M_{D}^{n+2}$, at the highest available 
centre-of-mass energy. For this reason, only the
HPC and the FEMC data recorded at 189~GeV 
were used to set a limit on the
gravitational scale. After the sensitivity had been optimised for
each calorimeter, the single photon sample consisted of
59 events with a photon in the HPC 
with $6<E_{\gamma}<50$~GeV and 45 events with a photon in the FEMC with 
$18<E_{\gamma}<50$~GeV. The numbers of events expected in 
the Standard Model were 64 and 41 for the two calorimeters respectively.
A cross-section limit of 
\begin{equation}
\sigma<0.24~\rm{pb}{\mbox{~~~at 95\% C.L.}}
\end{equation} 
results in limits on the fundamental mass scale of
$M_{D}>1.10$~TeV, $M_{D}>0.68$~TeV and $M_{D}>0.51$~TeV for 
2, 4 and 6 extra dimensions (Figure~\ref{grav}). This translates into
a limit on the size of the dimensions of $R<0.4$~mm for $n=2$.
If the systematic errors are taken into account,
the $M_{D}$-limit for two extra dimensions 
is reduced by 9\% and the limits for $n=$~4 and 6 by 3\%.

\begin{figure}[htb]
\centerline{\epsfig
{file=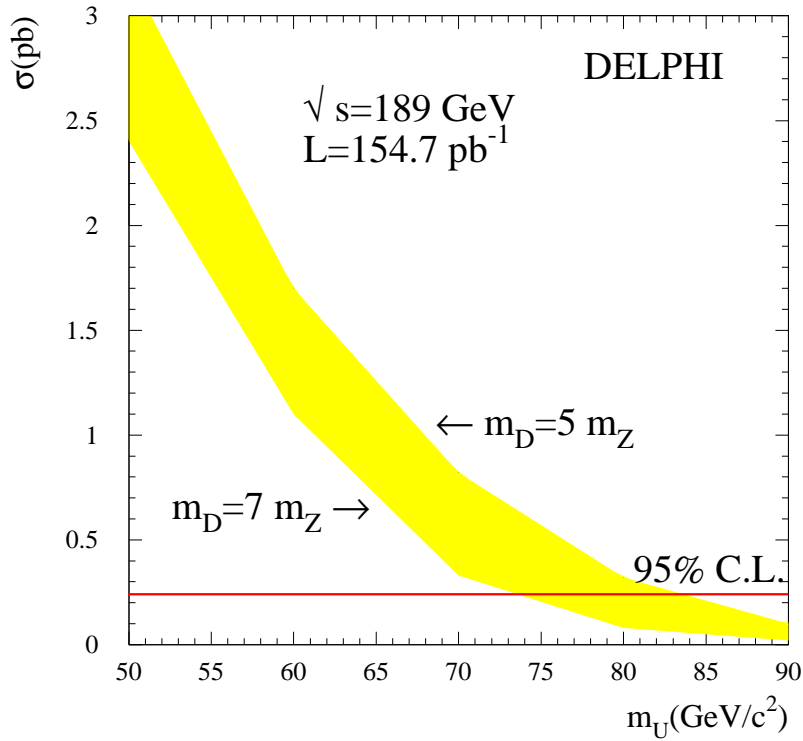,width=11.5cm,
bbllx=0pt,bblly=0pt,bburx=567pt,bbury=500pt}}
\caption[]{Cross section limit at 95$\%$~C.L. for the production of a $W$-type
$U$ boson from 189~GeV data. The shaded area shows the cross-section predicted
by the Preon Model described in the text.}
\label{preons} 
\end{figure}

\subsection{Limits on compositeness}

Composite models predict several new particles which do
not exist in the Standard Model.
A specific Preon Model is considered in this analysis~\cite{preons}.
This model considers leptons, quarks and weak bosons as composite particles.
Some of the predicted new particles contribute
to the cross-section of the process $e^+e^-\rightarrow\gamma+invisible~particles$.
At a relatively light mass scale, the model predicts the existence of objects
connected with neutrinos ($l_{S},\bar l_{S}$),
with down quarks ($q^{'}$) and with
$W$ bosons ($U^{\pm}$, $U^{0}$).
It also requires a new vector boson $D$, which could be 
several times more massive than the $Z^0$.
The $U^{0}$ boson decays invisibly and can be produced in the reaction 
$e^+e^- \rightarrow U^0 \bar{U^0}\gamma$,
contributing to the process $e^+e^-\rightarrow\gamma+invisible~particles$.
Also pairs of $l_{S} \bar l_{S}$ could be produced through 
$U^{\pm}$ exchange and contribute to the single-photon final state.

Calculating the cross-sections with the hypothesis that a composite
boson $D$ exists with mass between $m_{D}=5m_{Z^0}$ and 
$m_{D}=7m_{Z^0}$ and adding the contributions to the cross-sections coming from
direct production of $U^0\bar{U}^0$
pairs and the exchange of $U^{\pm}$, a limit can be obtained on
$m_U$ after subtracting the contribution expected from neutrino
production in the Standard Model.
The cross-section limit calculated from the HPC and the FEMC data was
$\sigma<0.24~\rm{pb}$ at 95\% C.L. as in the graviton analysis
and this translates into a limit on the $U$ boson mass which
ranges between $m_U>74-84$~GeV/c$^2$~at~95\% C.L. when
$m_{D}$ is varied in the range indicated above (Figure~\ref{preons}).
These limits are reduced by 4\% if the systematic errors are
taken into account.
Weaker limits have been determined at lower LEP2 energies~\cite{delphi_sg133}.
 
\subsection{Limit on the mass of the gravitino}

If the assumption is made that the gravitino is the lightest supersymmetric particle
(LSP), $\eeGG$ may be the only kinematically accessible supersymmetric process at LEP
as discussed and computed in~\cite{gravitino}. 
Lower limits on the mass 
of a light gravitino have been extracted in other LEP measurements~\cite{LEP_results},
at $p\bar{p}$ machines~\cite{pp}
and by using astrophysical constraints~\cite{astro} and $(g-2)_{\mu}$
measurements~\cite{g-2}.

To obtain a limit on the gravitino mass ($m_{\tilde{{G}}}$),
the radiative double 
differential cross-section $d^2\sigma/(dx_{\gamma},dcos\theta_{\gamma})$
given in~\cite{gravitino} for the radiative production ($\eeGG$),
was compared with the observed single photon data.
The largest sensitivity is obtained with photons at low energy and/or low polar angle. 
Single photon final states from the Standard Model process $\eenng$ have 
angular distributions similar to the signal, while the photon energy spectrum exhibits
the enhanced characteristic peak due to the radiative return to the $Z^0$, 
at $x_{\gamma}=1- \MZ^2/s$. Therefore, the optimal kinematic region in which to look for 
the signal is in the low photon energy region, well below the radiative return peak.
Since the signal cross-section grows as the sixth power of the centre-of-mass energy, 
the highest sensitivity is found at the highest beam energy.
For this reason, only the data taken at $\sqrt{s}=189$~GeV with the FEMC and the HPC
detectors have been used.
The different low energy regions available to the two calorimeters meant that
the HPC events dominated the measurement.
Combining the two calorimeters, the same limit of
$\sigma<0.24~\rm{pb}$ at 95\% C.L. was obtained as in the graviton analysis.
This corresponds to a lower limit on the gravitino mass which is
$$m_{\tilde{{G}}}> 10.0 \cdot 10^{-6} ~ \rm{eV/c^2}{\mbox{~~~at 95\% C.L.}}$$
Since the supersymmetry-breaking scale $|F|^{\frac{1}{2}}$ is related to the
gravitino mass by $|F|=\sqrt{\frac{3}{8\pi}/G_N}\cdot m_{\tilde{{G}}}$,
the limit on the scale is $|F|^{\frac{1}{2}}>204$~GeV.  
The effect of the systematic uncertainties on the $m_{\tilde{{G}}}$-limit is to 
lower it by 5\%. 

\begin{figure}[htb]
\centerline{\epsfig
{file=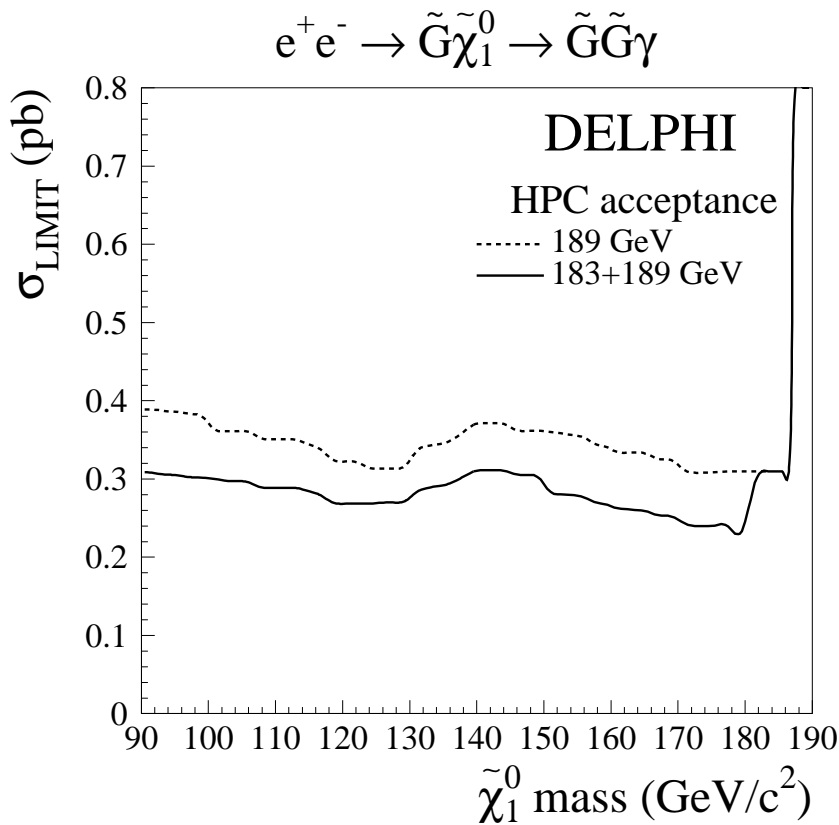,width=11cm}}
\caption[]{Upper limits for the cross-section of the process $\eeGGg$ at 95\%~C.L.
The dashed line shows the limit obtained with only the 189~GeV data while 
the full line represents the combined 183+189~GeV limit after scaling the low
energy data to 189~GeV (assuming the signal cross-section to scale as $1/s$).}
\label{chi_183} 
\end{figure}

\begin{figure}[hbt]
\begin{center}
 \mbox{\epsfxsize 7.8cm \epsfbox{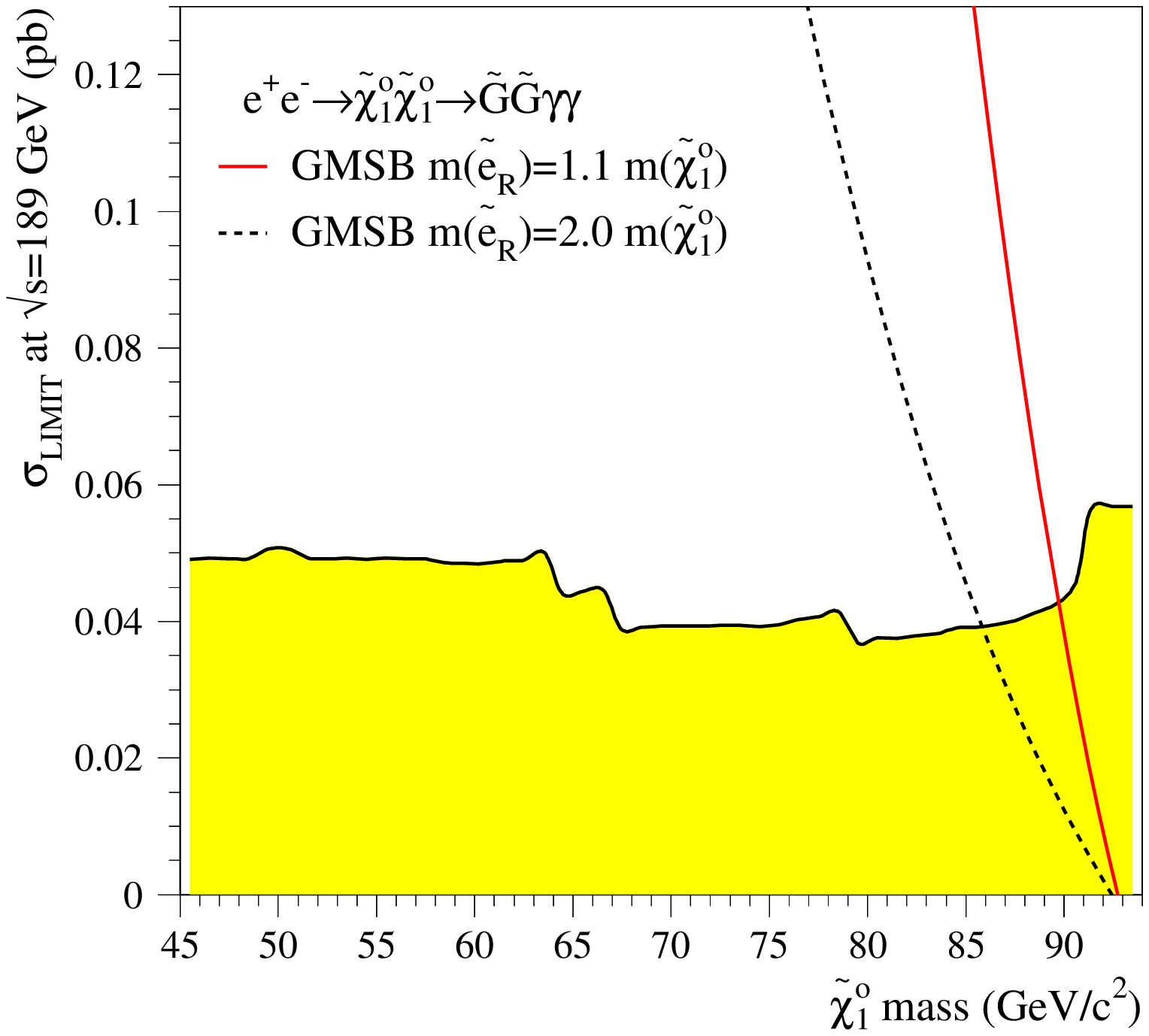}}
 \mbox{\epsfxsize 7.8cm \epsfbox{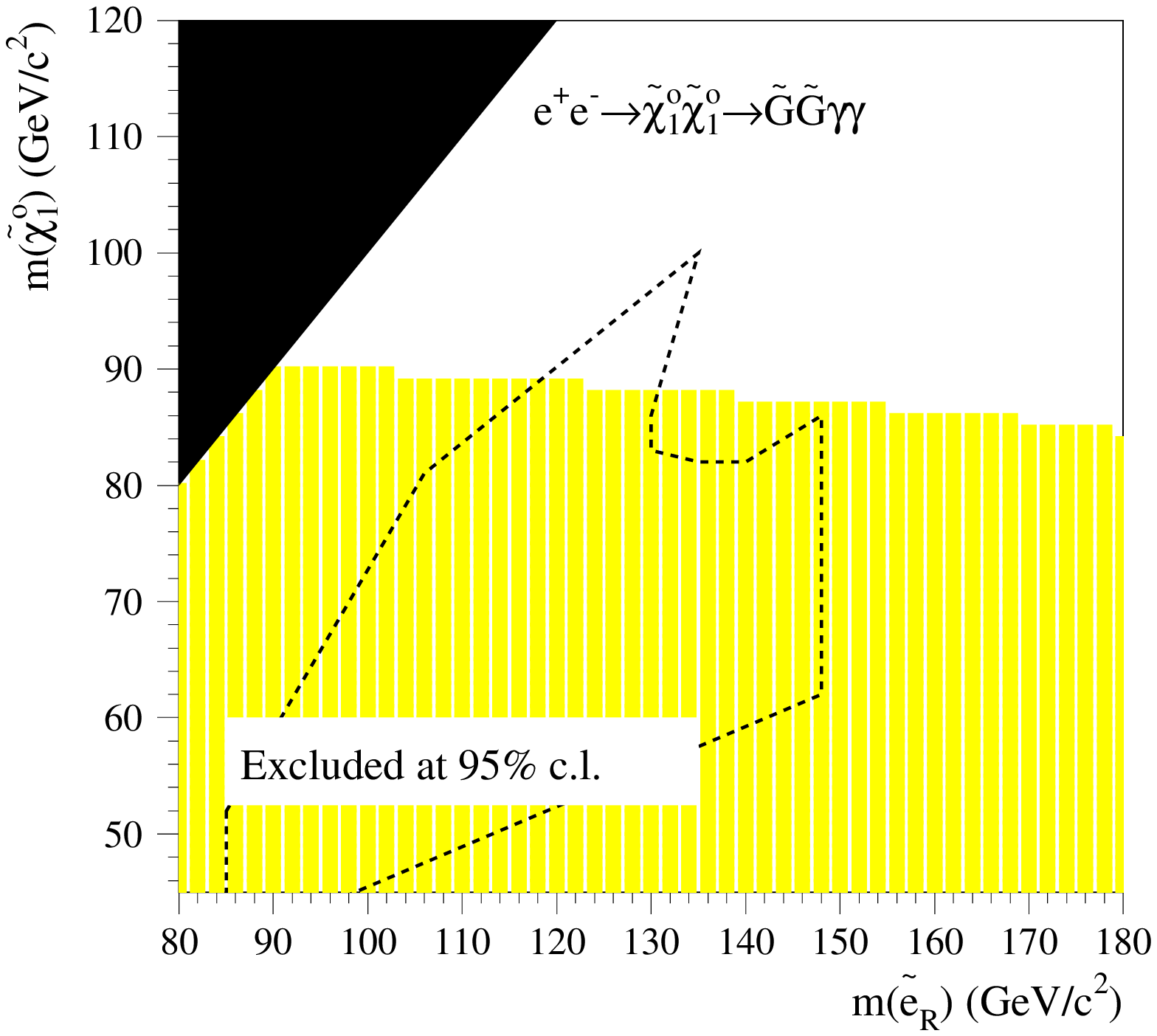}}
\end{center}
\caption[]{Left: Upper limit at 95\%~C.L. on the cross-section at $\sqrt{s}=$189~GeV of the process
$\eeGgGg$ as a function of the $\tilde{\chi}^0_1$ mass and the predicted
cross-section for two different assumptions for the selectron mass.
The limit was obtained by combining all data
taken at $\sqrt{s}=$130-189~GeV, assuming the signal cross-section scales as
$\beta/s$ (where $\beta$ is the neutralino velocity).
Right: The shaded area shows the
exclusion region in the $m_{\tilde{\chi}}$ versus $m_{\tilde{e}_R}$ plane,
calculated from the DELPHI data at $\sqrt{s}=$130-189~GeV.
The region compatible with the selectron interpretation~\cite{selectron} of the
CDF ee$\gamma\gamma$ event~\cite{CDF} is shown by the dashed line.} 
\label{chi_yyyy} 
\end{figure}

\begin{figure}[hbt]
\centerline{\epsfig {file=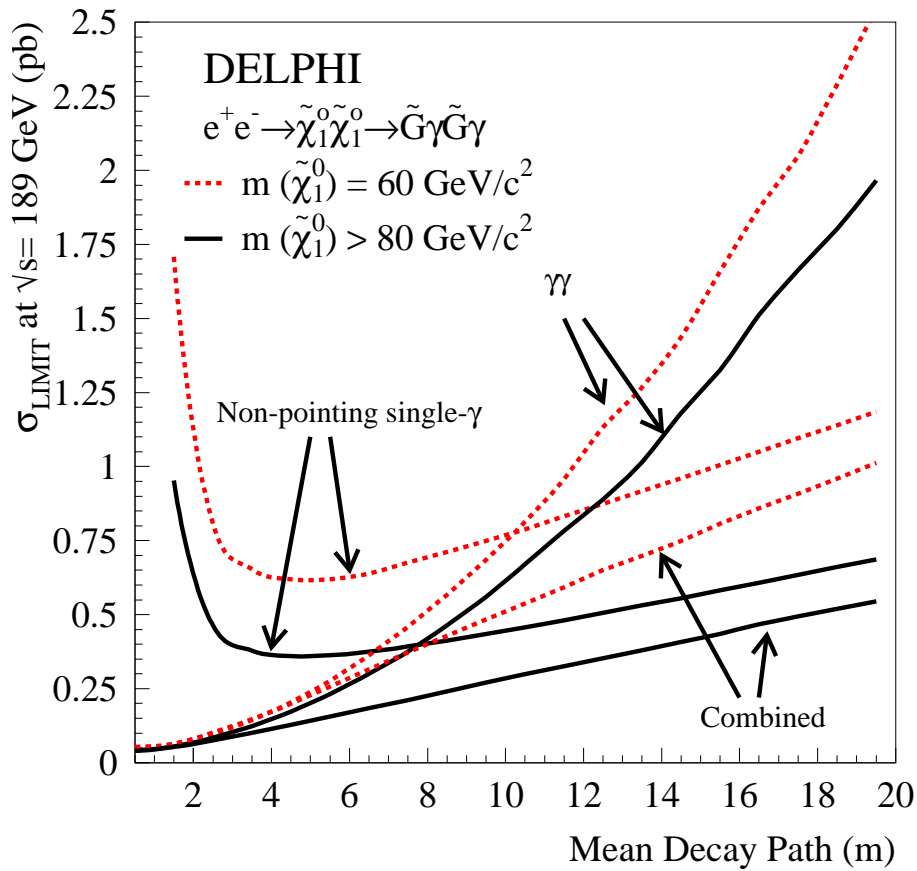,width=12.cm}}
\caption[]{Upper limit at 95\%~C.L. on the cross-section at $\sqrt{s}=$189~GeV
of the process $\eeGgGg$
as a function of the $\tilde{\chi}^0_1$ mean decay path for two
hypotheses for the neutralino mass: $m_{\tilde{\chi}}$~=~60~GeV/c$^2$ and
80~GeV/c$^2$$< m_{\tilde{\chi}}<\sqrt{s}/2$.} 
\label{chi_zz} 
\end{figure}

\begin{figure}[hbt]
\centerline{\epsfig {file=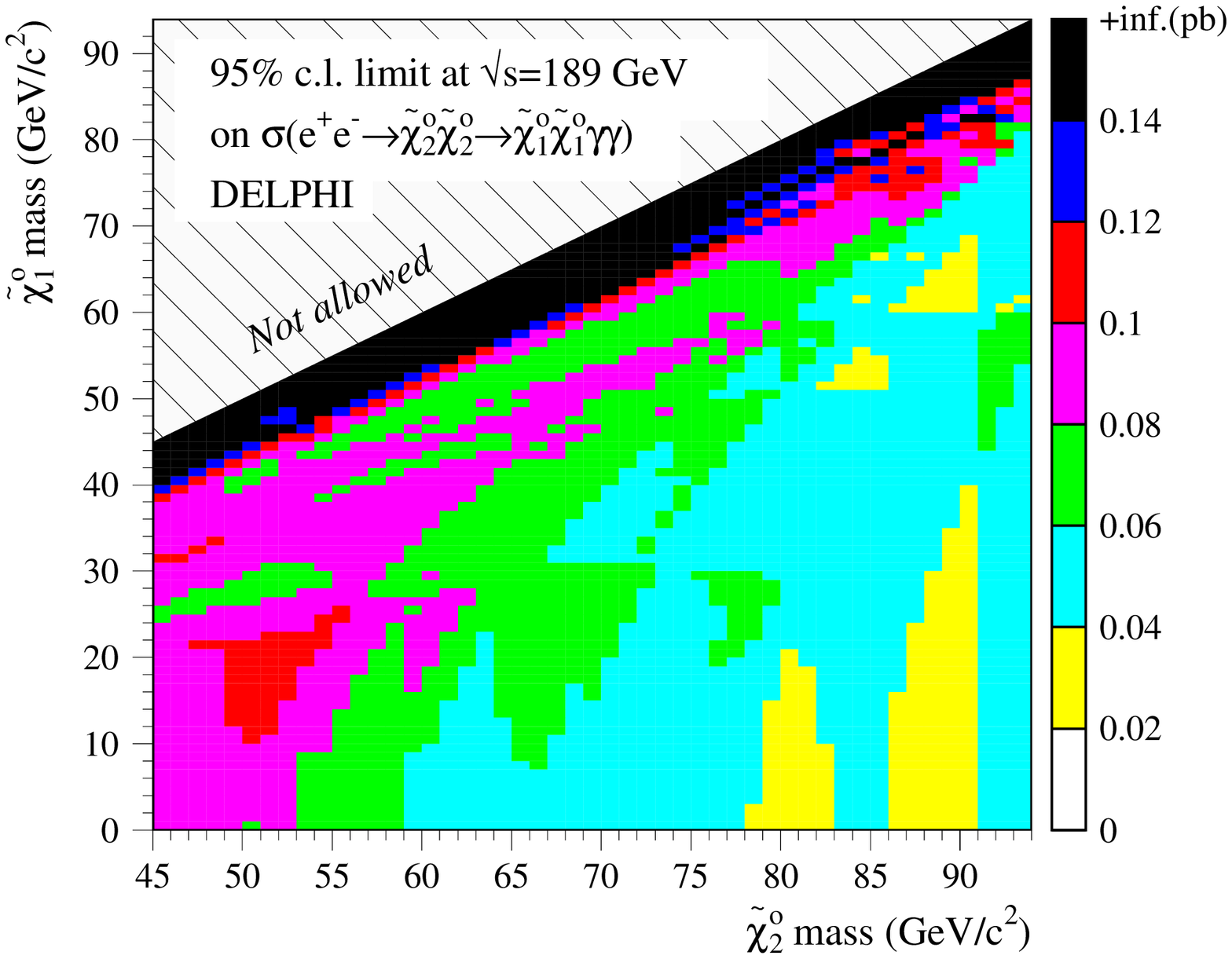,width=15.5cm}}
\caption[]{Upper limit at 95\%~C.L. on the cross-section at $\sqrt{s}=$189~GeV of the process
$\eeXgXg$ as a function of the $\tilde{\chi}^0_1$ and the $\tilde{\chi}^0_2$ mass.
The different shaded areas correspond to limits in pb as indicated by the shading scale on the 
right hand side. 
The limit was obtained by combining all data taken at $\sqrt{s}=$130-189~GeV, 
assuming the signal cross-section to scale as $\beta/s$.} 
\label{chi_yy} 
\end{figure}

\subsection{Limits on neutralino production if $\tilde{G}$ is the LSP}

Supersymmetric models such as the gauge-mediated supersymmetric (GMSB) model~\cite{GMSB}
or the ``no-scale" supergravity model (also known as the NLZ model)~\cite{NLZ}
predict that the gravitino $\tilde{G}$ is the lightest supersymmetric particle (LSP).
If the next lightest supersymmetric particle (NLSP) is 
the neutralino $\tilde{\chi}^0_1$, both single-photon and multi-photon production
can occur at LEP2 via the processes $\eeGGg$ and $\eeGgGg$.
While the rate of the former process is proportional to the inverse of the gravitino
mass squared, the di-photon process is independent of the gravitino mass.
Consequently, the single-photon process is expected to dominate only for
very light gravitinos and calculations done with the NLZ model at $\sqrt{s}=$~190~GeV
predict that $\eeGGg$ can be observed only if 
$m_{\tilde{G}} < 3\cdot10^{-5}$~eV/c$^2$~\cite{NLZ}. 

The cross-section limit for $\eeGGg$ was calculated from the energy distribution of
the expected events, generated with SUSYGEN~\cite{susygen}, and the observed single photon 
events in the angular region $45^\circ <\theta < 135^\circ$,
after taking into account the expected background from $\nu \bar{\nu} \gamma$.
The expected photon energy distribution from 
$\tilde{\chi}^0_1\rightarrow\tilde{G}\gamma$
is increasing with increasing neutralino mass ($m_{\tilde{\chi}}$) and the cut on 
$E_{\gamma}$ was changed with $m_{\tilde{\chi}}$ in such a way as to keep at least 90\% of 
the signal at all masses.
The resulting overall efficiency, including both the energy cut and the geometrical 
acceptance, varied 
between 55\% and 60\% for neutralino masses ranging from 50 to 180~GeV/c$^2$.
The calculated upper limit for the cross-section of the process $\eeGGg$
is given in Figure~\ref{chi_183} for 
the 189~GeV data alone and after combining the 183 and 189~GeV data
using a likelihood ratio method~\cite{LR}.
A branching ratio of 100\% for the process 
$\tilde{\chi}^0_1\rightarrow\tilde{G}\gamma$ was assumed.
The measured cross-section limit from the 183+189~GeV (189~GeV) data
corresponds to a limit on the neutralino mass 
of $m_{\tilde{\chi}^0_1}~\gt$~116~GeV/c$^2$ (110~GeV/c$^2$)
assuming $m_{\tilde{G}}=10^{-5}$~eV/c$^2$ 
and $m_{\tilde{e}}$~=150~GeV/c$^2$~\cite{NLZ}.

In the search for $\eeGgGg$ at $\sqrt{s}=$189~GeV, 5 
events were observed with 4.4 expected from 
$e^+e^-\rightarrow\nu\overline{\nu}\gamma\gamma(\gamma)$, which is the dominant
Standard Model background. This brings the total number of events found at
$\sqrt{s}=$130-189~GeV to 7 with 7.1 expected (Table~\ref{samples}).
Figure~\ref{chi_yyyy} shows the cross-section limit~\cite{LR} calculated from these events 
as a function of the $\tilde{\chi}^0_1$ mass (assuming a branching ratio of 100\% for 
$\tilde{\chi}^0_1\rightarrow\tilde{G}\gamma$) and 
the exclusion region in the $m_{\tilde{\chi}}$ versus $m_{\tilde{e}_R}$ plane.
The dependence of the signal cross-section on the selectron mass is due to the 
possibility of t-channel selectron exchange in the production mechanism.
As shown in Figure~\ref{chi_yyyy},
a lower limit of 86.0~GeV/c$^2$ (89.5~GeV/c$^2$)~at~95\% C.L. for the $\chi^0_1$ mass
can be deduced with the hypotheses $m_{\tilde{e}_R}=m_{\tilde{e}_L}=2m_{\tilde{\chi}}$
($m_{\tilde{e}_R}=m_{\tilde{e}_L}=1.1m_{\tilde{\chi}}$) and
 $\chi^0_1\approx\tilde B$.
In the extreme case $m_{\tilde{e}_L}\gg m_{\tilde{e}_R}$, 
the $\chi^0_1$ mass limit is reduced to 83.5~GeV/c$^2$
(88.5~GeV/c$^2$) at~95\% C.L.

If the gravitino mass is larger than 200-300~eV/c$^2$, the 
$\tilde{\chi}^0_1$ can have
such a long lifetime that it will decay far from the production point yet within
the detector. The signature for this case is
photons that do not point to the interaction region. 
If the decay length is long, the probability to detect both photons is small
and therefore single photon events were searched for which had a 
shower axis reconstructed in the HPC which gave a beam crossing point at 
least 40~cm away from the interaction point~\cite{delphi_gg}. Four events were found at
189~GeV with 5.2 expected, bringing the total at all energies to 6 with 7.9 
expected from Standard Model sources (Table~\ref{samples}).

Figure~\ref{chi_zz} shows the cross-section limit as a function of the
mean decay path of the neutralino using both the multi-photon events and the
non-pointing single photon events.

\subsection{Limits on neutralino production if $\tilde{\chi}^0_1$ is the LSP}

In other SUSY models~\cite{SUGRA} the
$\tilde{\chi}^0_1$ is the LSP and $\tilde{\chi}^0_2$ is
the NLSP. 
The $\eeXgXg$ process has an
experimental signature which is the same as for $\eeGgGg$
but with somewhat different kinematics due to the masses of the $\tilde{\chi}^0_1$
and $\tilde{\chi}^0_2$.
The previous DELPHI analysis at lower energies~\cite{delphi_gg} 
has now been repeated with the 189~GeV data sample.
Eight events remain after all cuts,
with 5.2 expected from the Standard Model background (Table~\ref{samples}). 
Figure~\ref{chi_yy} shows 
the cross-section limit calculated from the events collected at all energies
as a function of the $\tilde{\chi}^0_1$ and $\tilde{\chi}^0_2$
masses, assuming a branching ratio of 100\% for 
$\tilde{\chi}^0_2\rightarrow\tilde{\chi}^0_1\gamma$. 

\section{Conclusions}

With the 209~pb$^{-1}$ of data collected by DELPHI in 1997 and 1998 at
centre-of-mass energies of 183~GeV and 189~GeV, 
a study has been made of the production of events with a single 
photon in the final state
and no other visible particles.
Previous results on single non-pointing photons and on multi-photon 
final states have also been updated with 189~GeV data.

The measured single-photon cross-sections are in agreement with 
the expectations
from the Standard Model process $e^+e^- \rightarrow \nu \bar{\nu}\gamma$
and the number of light neutrino families is measured to be:
$$N_{\nu}=2.84\pm0.15(stat)\pm0.14(syst)$$

The absence of an excess of events with one or more photons in the final state
has been used to
set limits on the production of a new unknown model-independent
neutral state, a W-type $U$-boson as described by a composite
model, gravitons propagating in high-dimensional space,
a light gravitino and neutralinos.


\subsection*{Acknowledgements}
\vskip 3 mm
 We are greatly indebted to our technical 
collaborators, to the members of the CERN-SL Division for the excellent 
performance of the LEP collider, and to the funding agencies for their
support in building and operating the DELPHI detector.\\
We acknowledge in particular the support of \\
Austrian Federal Ministry of Science and Traffics, GZ 616.364/2-III/2a/98, \\
FNRS--FWO, Belgium,  \\
FINEP, CNPq, CAPES, FUJB and FAPERJ, Brazil, \\
Czech Ministry of Industry and Trade, GA CR 202/96/0450 and GA AVCR A1010521,\\
Danish Natural Research Council, \\
Commission of the European Communities (DG XII), \\
Direction des Sciences de la Mati$\grave{\mbox{\rm e}}$re, CEA, France, \\
Bundesministerium f$\ddot{\mbox{\rm u}}$r Bildung, Wissenschaft, Forschung 
und Technologie, Germany,\\
General Secretariat for Research and Technology, Greece, \\
National Science Foundation (NWO) and Foundation for Research on Matter (FOM),
The Netherlands, \\
Norwegian Research Council,  \\
State Committee for Scientific Research, Poland, 2P03B06015, 2P03B1116 and
SPUB/P03/178/98, \\
JNICT--Junta Nacional de Investiga\c{c}\~{a}o Cient\'{\i}fica 
e Tecnol$\acute{\mbox{\rm o}}$gica, Portugal, \\
Vedecka grantova agentura MS SR, Slovakia, Nr. 95/5195/134, \\
Ministry of Science and Technology of the Republic of Slovenia, \\
CICYT, Spain, AEN96--1661 and AEN96-1681,  \\
The Swedish Natural Science Research Council,      \\
Particle Physics and Astronomy Research Council, UK, \\
Department of Energy, USA, DE--FG02--94ER40817. \\



\end{document}